\documentclass[amssymb,amsmath,prd,twocolumn,superscriptaddress,nofootinbib]{revtex4-2}

\usepackage{graphicx}
\usepackage{bm}
\usepackage{amssymb,amsmath}
\usepackage{mathrsfs}
\usepackage{latexsym}
\usepackage{color}
\usepackage[normalem]{ulem}
\usepackage{dcolumn}
\usepackage[colorlinks=true,citecolor=blue,urlcolor=blue]{hyperref}
\usepackage[usenames,dvipsnames]{xcolor}
\usepackage{xspace}
\usepackage{graphicx}
\usepackage{multirow}
\usepackage{mathtools}
\usepackage{diagbox}

\allowdisplaybreaks

\usepackage[english]{babel}
\usepackage{xspace}
\usepackage[normalem]{ulem} 
\usepackage{color,soul}
\newcommand{\red}[1]{{\color{black}{#1}\xspace}}

\DeclareMathOperator{\dof}{dof}
\newcommand{\e}[1]{\times 10^{#1}}
\newcommand{\quant}[1]{\left(#1\right)}

\newcommand{\deriv}[2]{\frac{d#1}{d#2}}

\newcommand{\CIT}{\affiliation{Department of Physics, California Institute of Technology, Pasadena, California 91125, USA}}
\newcommand{\CITLab}{\affiliation{LIGO Laboratory, California Institute of Technology, Pasadena, California 91125, USA}}

\begin{document}
\newcommand{\npmixedlabel}{GP-mixed}
\newcommand{\nphadrlabel}{GP-hadronic}
\newcommand{\splabel}{SP}
\newcommand{\pplabel}{PP}
%

\newcommand{\nphadrilpostcost}{4.6 \e{-5}}
\newcommand{\nphadrilpsrcost}{5.3 \e{-4}}
\newcommand{\npmixilpostcost}{1.5 \e{-4}}
\newcommand{\npmixilpsrcost}{2.6 \e{-3}}
\newcommand{\ppilpostcost}{4.2 \e{-6}}
\newcommand{\ppilpsrcost}{4.1 \e{-5}}
\newcommand{\spilpostcost}{5.9 \e{-7}}
\newcommand{\spilpsrcost}{3.1 \e{-6}}

\newcommand{\nphadriqpostcost}{2.9 \e{-7}}
\newcommand{\nphadriqpsrcost}{6.2 \e{-7}}
\newcommand{\npmixiqpostcost}{5.0 \e{-7}}
\newcommand{\npmixiqpsrcost}{7.9 \e{-7}}
\newcommand{\ppiqpostcost}{2.7 \e{-7}}
\newcommand{\ppiqpsrcost}{1.3 \e{-6}}
\newcommand{\spiqpostcost}{8.4 \e{-8}}
\newcommand{\spiqpsrcost}{1.2 \e{-7}}

\newcommand{\nphadrqlpostcost}{6.9 \e{-4}}
\newcommand{\nphadrqlpsrcost}{8.9 \e{-3}}
\newcommand{\npmixqlpostcost}{2.0 \e{-3}}
\newcommand{\npmixqlpsrcost}{3.9 \e{-2}}
\newcommand{\ppqlpostcost}{2.1 \e{-4}}
\newcommand{\ppqlpsrcost}{4.2 \e{-3}}
\newcommand{\spqlpostcost}{3.3 \e{-5}}
\newcommand{\spqlpsrcost}{1.0 \e{-4}}

\newcommand{\nphadrbllowpostcost}{6.6 \e{-3}}
\newcommand{\nphadrblmedpostcost}{1.8 \e{-2}}
\newcommand{\nphadrblhipostcost}{9.1 \e{-3}}
\newcommand{\nphadrbllowpsrcost}{2.7 \e{-2}}
\newcommand{\nphadrblmedpsrcost}{1.0 \e{-1}}
\newcommand{\nphadrblhipsrcost}{8.5 \e{-2}}
\newcommand{\npmixbllowpostcost}{1.0 \e{-2}}
\newcommand{\npmixblmedpostcost}{5.2 \e{-2}}
\newcommand{\npmixblhipostcost}{4.9 \e{-2}}
\newcommand{\npmixbllowpsrcost}{5.5 \e{-2}}
\newcommand{\npmixblmedpsrcost}{2.6 \e{-1}}
\newcommand{\npmixblhipsrcost}{2.2 \e{-1}}
\newcommand{\ppbllowpostcost}{3.5 \e{-3}}
\newcommand{\ppblmedpostcost}{1.6 \e{-2}}
\newcommand{\ppblhipostcost}{1.1 \e{-2}}
\newcommand{\ppbllowpsrcost}{6.2 \e{-2}}
\newcommand{\ppblmedpsrcost}{2.2 \e{-1}}
\newcommand{\ppblhipsrcost}{9.6 \e{-2}}
\newcommand{\spbllowpostcost}{3.3 \e{-3}}
\newcommand{\spblmedpostcost}{6.5 \e{-3}}
\newcommand{\spblhipostcost}{2.0 \e{-3}}
\newcommand{\spbllowpsrcost}{6.7 \e{-3}}
\newcommand{\spblmedpsrcost}{1.0 \e{-2}}
\newcommand{\spblhipsrcost}{2.9 \e{-3}}

\newcommand{\nphadrclpostcost}{7.2 \e{-2}}
\newcommand{\nphadrclpsrcost}{2.2 \e{-1}}
\newcommand{\npmixclpostcost}{1.2 \e{-1}}
\newcommand{\npmixclpsrcost}{3.6 \e{-1}}
\newcommand{\ppclpostcost}{6.4 \e{-2}}
\newcommand{\ppclpsrcost}{4.7 \e{-1}}
\newcommand{\spclpostcost}{1.6 \e{-2}}
\newcommand{\spclpsrcost}{2.6 \e{-2}}

\newcommand{\nphadralphaccpostcost}{2.7 \e{-1}}
\newcommand{\nphadralphaccpsrcost}{1.8 \e{0}}
\newcommand{\npmixalphaccpostcost}{1.4 \e{0}}
\newcommand{\npmixalphaccpsrcost}{5.8 \e{0}}
\newcommand{\ppalphaccpostcost}{1.5 \e{-1}}
\newcommand{\ppalphaccpsrcost}{2.9 \e{-1}}
\newcommand{\spalphaccpostcost}{2.9 \e{-2}}
\newcommand{\spalphaccpsrcost}{7.2 \e{-2}}

\newcommand{\sprlambdatpostcost}{2.5 \e{-2}}
\newcommand{\sprlambdatpsrcost}{1.8 \e{-2}}
\newcommand{\nphadrrlambdatpostcost}{1.2\e{-1}}
\newcommand{\nphadrrlambdatpsrcost}{1.3 \e{-1}}
\newcommand{\npmixrlambdatpostcost}{1.4 \e{-1}}
\newcommand{\npmixrlambdatpsrcost}{1.7 \e{-1}}
\newcommand{\pprlambdatpostcost}{7.7 \e{-2}}
\newcommand{\pprlambdatpsrcost}{1.4 \e{-1}}

\title{Assessing equation of state-independent relations for neutron stars with nonparametric models}

\author{Isaac Legred}
\email{ilegred@caltech.edu}
\CIT \CITLab 

\author{Bubakar O. Sy-Garcia}
\email{sygarcia@wisc.edu}
\affiliation{Department of Astronomy, University of Wisconsin-Madison,
Madison, Wisconsin 53706 USA }

\author{Katerina Chatziioannou} 
\email{kchatziioannou@caltech.edu} 
\CIT \CITLab

\author{Reed Essick}
\email{essick@cita.utoronto.ca}
\affiliation{Canadian Institute for Theoretical Astrophysics, University of Toronto, Toronto, ON M5S 3H8}
\date{\today}

\begin{abstract}
    Relations between neutron star properties that do not depend on the nuclear equation of state offer insights on neutron star physics and have practical applications in data analysis. Such relations are obtained by fitting to a range of phenomenological or nuclear physics equation of state models, each of which may have varying degrees of accuracy. In this study we revisit commonly-used relations and re-assess them with a very flexible set of phenomenological nonparametric equation of state models that are based on Gaussian Processes. Our models correspond to two sets: equations of state which mimic hadronic models,   and equations of state with rapidly changing behavior that resemble phase transitions. 
    We quantify the accuracy of relations under both sets and discuss their applicability with respect to expected upcoming statistical uncertainties of astrophysical observations. We further propose a goodness-of-fit metric which provides an estimate for the systematic error introduced by using the relation to model a certain equation-of-state set.
    Overall, the nonparametric distribution is more poorly fit with existing relations, with the I--Love--Q relations retaining the highest degree of universality.  Fits degrade for relations involving the tidal deformability, such as the Binary-Love and compactness-Love relations, and when introducing phase transition phenomenology. For most relations, systematic errors are comparable to current statistical uncertainties under the nonparametric equation of state distributions.   
\end{abstract}

\maketitle

\section{Introduction}
\label{sec:introduction}

While most neutron star (NS) properties depend sensitively on the unknown equation of state (EoS) of dense nuclear matter, some properties are interrelated in an approximate EoS-independent way~\cite{Yagi:2016bkt}. The impact of EoS-independent relations ranges from enhancing our understanding of NS physics~\cite{Lattimer:2000nx,Yagi:2014qua,Yagi:2014bxa,Bauswein:2015yca,Saes:2021fzr} to practical applications in analyses of data. For example, relations between the NS multiple moments~\cite{Pappas:2013naa,Yagi:2014bxa,Stein:2013ofa,Chatziioannou:2014tha} have led to a generalization of the no-hair theorem for black holes to the three-hair relations for Newtonian NSs~\cite{Stein:2013ofa}, while the so-called ``I--Love--Q" relations~\cite{Yagi:2013bca,Yagi:2013awa} have been attributed to the self-similarity of isodensity contours~\cite{Yagi:2014qua}.
On the data analysis side, EoS-independent relations reduce the number of degrees of freedom~\cite{Yagi:2016qmr,Chatziioannou:2018vzf,Zhao:2018nyf,Chatziioannou:2020pqz,Xie:2022brn} and enable consistency tests~\cite{Yagi:2013awa,Samajdar:2020xrd,Silva:2020acr,Wijngaarden:2022sah,Breschi:2023mdj}.

EoS-independent relations may include static or dynamic and macroscopic or microscopic quantities. One of the earliest proposed such relation is the one between the (complex) NS modes and their mass and radius, which can be used to translate gravitational wave (GW) observations from isolated NSs to constraints on the radius~\cite{Andersson:1997rn,Andersson:1996pn,Tsui:2004qd,Benhar:2004xg,Lau:2009bu}.
Additionally, relations including the NS tidal parameters can simplify analysis of GW data. In general, the signal emitted during the coalescence of two NSs depends on a list of tidal deformability parameters and the rotational quadrupole moment of each star. Relations between the different tidal parameters and the quadrupole moment~\cite{Yagi:2013sva,Yagi:2016qmr,Yagi:2013bca,Yagi:2013awa} reduce the number of free parameters to one per star, typically the so-called dimensionless tidal deformability $\Lambda_i$, $i\in \{1,2\}$. A relation between $\Lambda_1$ and $\Lambda_2$ (and the binary mass ratio) further reduces the number of free parameters to just one~\cite{Yagi:2015pkc,Chatziioannou:2018vzf,Abbott:2018exr, Raithel:2018ncd, De:2018uhw, Benitez:2020fup}. 

EoS-independent relations are typically constructed empirically by fitting a large number of EoS models, obtained either through phenomenological or theoretical nuclear models. Their applicability is therefore limited to the nuclear physics represented in the set of EoSs, while deviations from the relations may be a sign of new (relevant) physics. For example, an observed deviation from the relation between the frequency content of the post-merger GW signal from a NS coalescence and the tidal properties of the pre-merger signal that hold for hadronic matter~\cite{Bauswein:2011tp,Takami:2014zpa,Bauswein:2015yca,Bernuzzi:2015rla,Foucart:2015gaa,Lehner:2016lxy} can signal the presence of quark matter in the merger remnant~\cite{Bauswein:2018bma,Most:2018eaw,Wijngaarden:2022sah,Breschi:2023mdj}. The breakdown of universal behavior in a catalog of observations can further be used to identify outliers that can be attributed to quark matter~\cite{Chatziioannou:2019yko} or NS-black hole binaries~\cite{Chen:2019aiw}.
 
Beyond relations breaking down outside their regime of validity, EoS-independent relations display different degrees of independence even within it, which furthermore varies across the NS parameter space. The set of EoSs the relation is fitted to sensitively impacts the degree of EoS-independence. A potential choice of such a set is EoS candidates from nuclear theory, and corresponds to evaluating the degree of independence present in existing nuclear models~\cite{Yagi:2013bca,Yagi:2015pkc,Papigkiotis:2023onn}. Nonetheless, the extent to which current nuclear models cover the entire range of possible behaviors of matter at high densities is unclear.  

More extended sets of EoSs can be obtained by considering phenomenological models, designed to mimic nuclear theory while maintaining some degree of flexibility at high densities. Examples of such phenomenological models include piecewise polytropes~\cite{Read:2008iy,OBoyle:2020qvf} and spectral representations~\cite{Lindblom:2010bb,2012PhRvD..86h4003L,Lindblom:2013kra}. This approach leads to large sets of EoSs and statistical distributions on the EoS which can further be conditioned on astrophysical observations. Such studies directly quantify the impact of astrophysical constraints on the degree of EoS-independence compared to fully agnostic nuclear behavior. For example,~\citet{Carson:2019rjx} considered spectral EoSs that have been conditioned on GW170817~\cite{TheLIGOScientific:2017qsa,Abbott:2018wiz} and found that the degree of EoS-independence can be improved by more than 50$\%$ compared to an agnostic EoS set. Similar improvements have been reported in~\cite{Godzieba:2021vnz,Godzieba:2020bbz,Nath:2023ela}.   

\begin{figure}
    \centering
    \includegraphics[width=.49\textwidth]{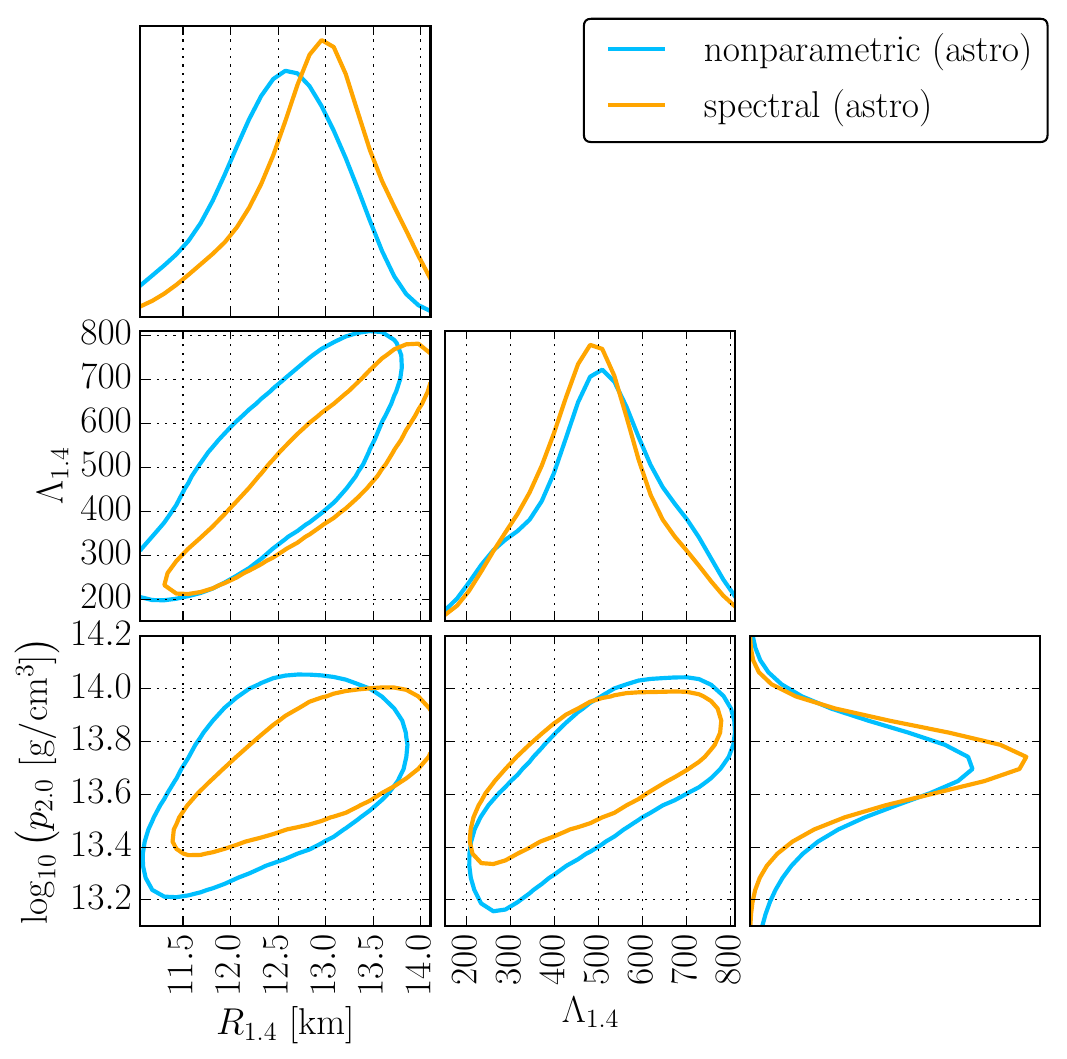}
    \caption{The astrophysically-informed posterior distributions for $R_{1.4}$, $\Lambda_{1.4}$, and $p_{2.0}$ when using nonparametric (blue) and spectral (orange) EoSs. Astrophysical distributions are conditioned on pulsar mass, mass-radius, and mass-tidal deformability measurements; see Sec.~\ref{sec:EoSset}.  The spectral EoS result shows less variability in $R_{1.4}$ at a fixed value of $\Lambda_{1.4}$ than the nonparametric one. This suggests that the degree of EoS-independence in $R_{1.4}$--$\Lambda_{1.4}$ is linked to the flexibility of the EoS model. Similar conclusions hold for $p_{2.0}$.}
    \label{fig:R14L14posterior}
\end{figure}

Though more generic than a set of selected nuclear models, parametric EoS representations are still limited in flexibility by the functional form of the EoS, which is usually not determined from first principles. This can lead to strong correlations between the EoS at different densities that are not an outcome of nuclear insight, but of the arbitrary functional form of the representation~\cite{Legred:2022pyp}. These correlations effectively cause many EoSs in the fitting set to share similar macroscopic and microscopic features, mimicking or strengthening true EoS-independence~\cite{Legred:2022pyp}. 
Figure~\ref{fig:R14L14posterior} shows an example of such emerging EoS-independence in the radius $R_{1.4}$ and dimensionless tidal deformability $\Lambda_{1.4}$ of a $1.4M_{\odot}$ NS, and the pressure at twice saturation\footnote{We define the saturation density as $2.8\times 10^{14}\,\mathrm{g}/\mathrm{cm}^3$.} $p_{2.0}$. The $R_{1.4}$--$\Lambda_{1.4}$ relation is an outcome of the so called C--Love relation~\cite{Maselli:2013mva,Yagi:2016bkt} (discussed more later), while a correlation with $p_{2.0}$ has been observed in several theoretical models~\cite{Lattimer:2000nx}, \red{and is analogous to the C--$\alpha_c$ relation described later}.   Using the spectral parameterization, perfect knowledge of $\Lambda_{1.4}$ would give a $R_{1.4}$ uncertainty of $\sim 1$\,km, consistent with the error in the C--Love relation computed in~\cite{Carson:2019rjx}.\footnote{\red{When computing the compactness (and throughout unless otherwise stated), we use units with  $G = c = 1$}.}

Figure~\ref{fig:R14L14posterior} also shows the same relations obtained with a more flexible set of nonparametric EoSs based on Gaussian Processes~\cite{Landry:2018prl, Essick:2019ldf} that is only minimally informed by nuclear physics. The nonparametric EoSs are drawn from a collection of Gaussian Processes and explore a wide rage of intra-density correlations lengths and strengths. As shown in~\citet{Legred:2022pyp}, this EoS set is extremely agnostic and intra-density correlations are only imposed by physical considerations such as causality and thermodynamic stability. Due to its flexibility, the set also inherently includes EoSs with phase-transition-like behavior, including nonmonotonic behavior in the speed of sound and multiple stable branches~\cite{Essick:2023fso}. As expected, under the nonparametric EoSs, perfect knowledge of $\Lambda_{1.4}$ yields an increased uncertainty in $R_{1.4}$ of $\sim2$\,km, larger than the nominal error of the $C$-Love relation.    

In this work and motivated by Fig.~\ref{fig:R14L14posterior}, we revisit common EoS-independent relations and assess them under nonparametric EoSs.  Following Ref.~\cite{Carson:2019rjx}, we evaluate EoS-independent relations separately against hadronic EoS sets as well as mixed hadronic and hybrid EoSs. 
Because of the difficulty in fitting the relation over an unstable branch of the $M$--$R$ relation,  we only study EoSs with a single stable branch, thus restricting to weak phase transitions.  
We also consider EoSs that are only required to be consistent with the existence of massive pulsar measurements, contrasted with a set required to be consistent with additional GW and X-ray measurements~\cite{Legred:2021hdx}. 

With a focus on the applicability of EoS-independent relations, we further revisit the issue of EoS-independence across the parameter space. \red{In general, relations are most useful in the regions of parameter space where data are most informative, since tight constraints on some parameters can be interpreted as constraints on other parameters}. A higher degree of EoS-independence in these regions will therefore expand their applicability. For example, the relations that link the dimensional tidal deformabilities of two NSs in a binary to each other are most useful for NS with masses $\lesssim 1.7M_{\odot}$ as GW observations are largely uninformative about the tidal properties of more massive NSs~\cite{Abbott:2020uma,Chen:2020fzm}. In Sec.~\ref{sec:goodness-of-fit}, we propose a statistic to measure the goodness-of-fit of an EoS-independent relation, by comparing to a \emph{tolerance factor} which is chosen based on the application.  Fitting via optimization of this metric allows more control over the precision of the EoS-independent relation as a function of NS mass.

With the extended EoS set and goodness-of-fit metric in hand, we revisit the following relations in Sec.~\ref{sec:results} :

\begin{itemize}
    \item \emph{I--Love--Q}~\cite{Yagi:2013bca,Yagi:2013awa}, Sec.~\ref{sec:results-I--Love--Q}: a relation between the (normalized) moment of inertia $I$, the tidal deformability $\Lambda$, and the rotational quadrupole moment $Q$ of a NS.  The I--Love--Q relations remain highly universal, likely useful even with sensitivies more than ten times current GW detectors.

    \item \emph{C--Love}~\cite{Maselli:2013mva,Yagi:2016bkt}, Sec.~\ref{sec:binary-love}: a relation between the compactness $C=m/R$ and the tidal deformability $\Lambda$ of a NS. Its main applicability is in translating GW tidal constraints to radii, given the NS mass $m$. The C--Love relation is relatively non-universal; for nonparametric EoS distributions, it leads to systematic errors of $\sim30\%$ compared to statistical uncertainties at current sensitivity.  This holds true for EoSs both with and without strong phase transitions.

    \item \emph{Binary-Love}~\cite{Yagi:2015pkc}, Sec.~\ref{sec:results-c-lambda}: a relation between the dimensionless tidal deformabilities of two NSs in a binary $\Lambda_1$ and $\Lambda_2$ given the mass ratio $q$. Its main applicability is in reducing the number of parameters in GW analyses on NS binaries, though its EoS-independence breaks down for EoSs with phase transitions~\cite{Tan:2021nat,Carson:2019rjx}. The binary-Love relation is similarly non-universal under the nonparametric EoS distribution with systematic errors $\sim50\%$ of current statistical uncertainties. The Binary-Love relation universality is further degraded for EoSs with phase transitions.
    \item \emph{$R_{1.4}$-Love}~\cite{Raithel:2018ncd, De:2018uhw, Zhao:2018nyf}, Sec.~\ref{sec:results-R-Love}: a relation between the NS radius and the chirp mass and chirp tidal deformability of a NS binary, essentially combining the C--Love and Binary-Love relations above.  $R_{1.4}$-Love likely would introduce bias before the advent of next-generation detectors, with systematic errors becoming comparable to statistical uncertainties for a GW170817-like but $\mathcal O (3-5)$ times louder.
    \item \emph{$\alpha_c$-C}~\cite{Saes:2021fzr}, Sec.~\ref{sec:results-alphac-c}: a relation between the EoS stiffness measure $\alpha_c\equiv p_c/\epsilon_c$ where $p_c$ and $\epsilon_c$ are the central pressure and energy density respectively, and the compactness. The $\alpha_c-C$ relation is a very poor fit to the nonparametric mixed distribution with systematic errors greater than or equal to current statistical uncertainties. The relation is somewhat better fit by the parametric and hadronic nonparametric distributions.  
\end{itemize}
%

\section{Goodness-Of-Fit and Quantifying EoS-independence}
\label{sec:goodness-of-fit}

In this section we formalize the discussion of EoS-independent relations by quantifying EoS-independence through a goodness-of-fit metric in Sec.~\ref{sec:statistic-definition}, introducing a tolerance factor for the fit in Sec.~\ref{sec:tolerance}, and describing the EoS sets we use in Sec.~\ref{sec:EoSset}.

In general, an individual NS is characterized by an EoS $\epsilon$ and the NS central density $\rho_c$.  Given two NS properties $F(\epsilon, \rho_c)$ and $G(\epsilon, \rho_c)$ which are each one-to-one\footnote{If $F$ is not one-to-one with $\rho_c$ (for example the mass $m(\epsilon, \rho_c)$ for EoSs with multiple stable branches and twin stars), then this construction works on each monotonic branch.} with $\rho_c$, we define their relation $G(\epsilon, F(\epsilon, \rho_c))$.  Remarkably, for a number of property pairs the induced function $G(\epsilon, F(\epsilon, \rho_c))$ is nearly independent of $\epsilon$.  These are so-called universal, or EoS-independent relations.

\subsection{Defining a goodness-of-fit metric}
\label{sec:statistic-definition}

Following~\cite{Yagi:2013awa}, we fit an analytic phenomenological approximant to the EoS-independent relation
\begin{equation}
    \tilde G(F;\bm{\theta}) \approx G(\bullet, F(\bullet, \rho_c))\,,
\end{equation}
where $\bullet$ in place of the EoS $\epsilon$ indicates this should hold regardless of the EoS and $\bm{\theta}$ are fitting parameters.  
Given a functional form for $\tilde G(F;\bm{\theta})$ (typically in terms of simple functions such as polynomials and logarithms) and a particular EoS $\epsilon$, we select $\bm{\theta}$ such that a goodness-of-fit metric is minimized.  A least-squares metric is\footnote{Though this metric is not strictly a $\chi^2$ statistic, as there is no statistical interpretation of the scatter which induces the $\chi^2$, we use familiar notation since many conventional intuitions hold.  For instance, $\chi^2/N_{\rm dof} =1$ is a threshold for a good fit, and any value significantly smaller than 1 would be regarded as overfitting~\cite{numericalrecipes}.  In our case, we expect the EoS-independent relations to overfit the ``data", $\chi^2/N_{\rm dof} \ll 1$. A large value would be considered a poor fit.}
\begin{equation}
    \label{eq:chi-square-eos}
    \chi^2_\epsilon(\bm{\theta}) \equiv \sum_i^N \frac{\left[\tilde G(F_i;\bm{\theta}) -  G(\epsilon, F_i(\epsilon, \rho_{c,i}))\right]^2}{\sigma_i^2}\,,
\end{equation}
where $i$ iterates over individual stellar solutions (i.e., central densities),  $\sigma_i$ is a tolerance factor for the goodness-of-fit of data point $i$, and $N$ represents the number of central densities the relations are evaluated at. In what follows we use $N=200$ which ensures smooth relations and that $\chi^2/(N-N_p)$, the $\chi^2$ per number of degrees of freedom (with $N_p$ the number of parameters of the fit), is independent of $N$.

Unless otherwise stated, we fit each relation on a grid of NS central densities.  We build a linear grid for each EoS in the central rest-mass density, $\rho_c$, for $1.0M_{\odot}$ to $M_{\max}$, the maximum TOV mass.  We use only EoSs with a single stable branch in the $M-R$ relation.  Both the choice of grid used, and the truncation are inputs and represent a \emph{de facto} choice of relative significance weighting between mass scales, which may or may not be realistic depending on the true distribution of NS masses (equivalently, given an EoS, the distribution  of central densities).  For most EoSs, the spacing of central density favors higher masses; given the uncertainty in populations of NSs, we do not attempt to modify this distribution substantially.  Implications of this choice are discussed further in Sec.~\ref{sec:discussion}.

The tolerance factor $\sigma_i$ can be freely chosen and, as its name suggests, quantifies the degree of deviation from EoS-independence we tolerate. Different choices for $\sigma_i$ will result in different best-fit $\bm{\theta}$ parameters and goodness-of-fit estimates. We discuss the tolerance factor extensively in the next section.

Beyond a single EoS $\epsilon$, we consider a (normalized) distribution on EoSs $P(\epsilon)$, potentially conditioned on observations. The distribution-dependent goodness-of-fit is then defined as the distribution average of $\chi^2_{\epsilon}$ over $P(\epsilon)$,
\begin{equation}
    \chi^2(\bm{\theta}) \equiv \int \chi^2_{\epsilon}(\bm{\theta}) P(\epsilon) d\epsilon =  \sum_{\epsilon} P_{\epsilon} \chi^2_{\epsilon}(\bm{\theta})\,,
    \label{eq:chi-square-total}
\end{equation}
where $P_{\epsilon}$ is the weight of each EoS in the distribution $P(\epsilon)$.  EoSs are sampled for the Monte Carlo sum by directly sampling an EoS prior set for each distribution; we use the same prior distributions as~\cite{Legred:2022pyp}.
In Eq.~\eqref{eq:chi-square-total} the fitting parameters $\bm{\theta}$ are shared among and fitted with all $\chi^2_{\epsilon}(\bm{\theta})$ --this is equivalent to seeking a set of parameters which are EoS-independent over $P(\epsilon)$.  In practice, we sample EoSs uniformly  from the approximate support of $P_{\epsilon}$, i.e. $\{\epsilon | P_{\epsilon}>P_{th}\} $ for some threshold $P_{th}$, and weigh each EoS draw  by $P_{\epsilon}$.  This allows us to better resolve the ``tails" of the EoS distribution where $\chi_{\epsilon}^2$ may be large. W sample 1000 draws from the  given EoS set in order to approximate the integral, as we found reasonable convergence of the total $\chi^2$ was achieved by this point for all EoS distributions (see Sec.~\ref{sec:EoSset}).

\subsection{Role of the tolerance factor}
\label{sec:tolerance}

Setting $\sigma_i=1$ would be sufficient to uniquely specify a fitting problem for $\bm{\theta}$ if the goal is simply to obtain a fit.  However, in this case, no information about goodness-of-fit is contained in Eq.~\eqref{eq:chi-square-total}, because rescaling $\sigma_i \to \alpha \sigma_i$ changes $\chi^2 \to \chi^2/\alpha^2$; any level of goodness-of-fit could be achieved by rescaling. In fact, no specific fit corresponds in any sense to the ``best fit" possible as a different (non-constant) $\sigma_i$ would produce a different fit. This is analogous to a nonlinear change of variables producing a different fit.   We instead select $\sigma_i$ by considering the tolerance we have for error in the EoS-independent relation. This results in a $\chi^2(\bm{\theta})$ that is simultaneously used during the fitting procedure and whose (dimensionless) numerical value can be interpreted as a goodness-of-fit.  

To clarify, we dig further into a common application of EoS-independent relations in inference, namely the computation of certain NS properties from others without knowledge of the EoS.  The Binary-Love relations~\cite{Yagi:2015pkc, Yagi:2016bkt} facilitate the computation of the tidal deformability of one NS $\Lambda_1$ from that of another $\Lambda_2$ given their mass ratio $q$~\cite{Chatziioannou:2018vzf}. The systematic error in the estimation of $\Lambda_1$ due to the relation's error is $\delta \Lambda_{sys}$. Whether this systematic error is tolerable in a GW analysis depends on the statistical measurement uncertainty $\delta \Lambda_{stat}$. If $\delta \Lambda_{sys} \gtrsim \delta \Lambda_{stat}$, then the application of the relation introduces an uncertainty comparable to the statistical uncertainty, which is undesirable.  If, however, $\delta \Lambda_{sys} \ll \delta \Lambda_{stat}$, then the relation may be useful as the statistical uncertainty dominates.  This consideration motivates choosing the tolerance factor $\sigma_i$ to be the approximate measurability of the quantity of interest. In doing do, the goodness-of-fit $\chi^2(\bm{\theta})$ is a direct check of the relation between $\delta \Lambda_{sys}$ and  $\delta \Lambda_{stat}$.  Unless otherwise stated, throughout this work we use a fiducial estimate of $\delta \Lambda_{stat} = 210$,  a constant motivated by the tidal measurement of GW170817 and rescaling the symmetric $90\%$ region to 1-$\sigma$~\cite{Abbott:2018exr}.  Improvements in detector sensitivity mean that a GW170817-like event observed today would have a lower statistical uncertainty; per Eq.~\eqref{eq:chi-square-eos}, halving the statistical uncertainty in $\Lambda$ would increase the $\chi^2$ (i.e., decrease the goodness-of-fit) by a factor of $4$. 

In certain cases the measurability of NS tidal deformability is a very poor estimate for the measurability of other NS properties.  
For example, for higher-mass NSs, the compactness will likely be better measured by non-GW techniques, such as X-ray pulse-profile modeling.  
In such cases, we approximate statistical uncertainty by assuming that the compactness, $C(M)$, can be measured to within $\delta C = 0.02$, a constant representing the uncertainty from X-ray observations~\cite{Bogdanov:2021yip,Riley:2019yda, Riley:2021pdl, Miller:2019cac, Miller:2021qha}.  See Secs.~\ref{sec:results-c-lambda} and~\ref{sec:results-alphac-c} for more details of how we simultaneously incorporate separate estimates of NS measurability.

Generically, the $\chi^2$ value represents how poorly fit the relation is to the EoS distribution.  Per Eq.~\eqref{eq:chi-square-eos}, $\chi^2_{\epsilon}$ represents the square error in the quantity predicted by the relation relative to the tolerance factor.  Given a value for $\chi^2$, the typical error in the underlying variable is 
\begin{equation}
\Delta G \sim \sigma(F) \sqrt{\chi^2/N_{\rm dof}}\,.
\end{equation}
Here, $\sigma (F)$ represents the tolerance factor on the quantity $F$ used in evaluating the fit. 
This is to be taken as an order of magnitude estimate, and is useful for quickly diagnosing the error expected from applying an EoS-independent relation.  For example, if $\sigma(F)$ represents statistical measurement uncertainty, then a $\chi^2$ value of $10^{-4}$ indicates that systematic errors in parameters are of order $0.01 = 1\%$ statistical uncertainties.  

An alternative choice for the tolerance factor would be 
\begin{equation}
    \sigma(F) = G(F)\,.
\end{equation}
This corresponds to constant tolerance for the \emph{fractional} error in the fit.  This tolerance factor is independent of measurement uncertainty, and so the best fit bears a different interpretation.    In many cases a constant relative tolerance may be preferred, especially when an observable varies over orders of magnitude.   We give an example of a fit where a constant fractional error tolerance gives a seemingly better fit in Sec.~\ref{sec:binary-love-tolerance-change}. The $\chi^2$ in this case is a measure of the total fractional deviation in the relation.

Nonetheless, there are subtleties to interpretation of the $\chi^2$ value within the fractional uncertainty approach.   For example, assume we decided to try to identify EoS-independent relations for $R(M)$ or $\Lambda(M)$. Since $R(M)$ is approximately constant for a large class of EoSs, we adopt a constant fit : 
\begin{equation}
    \chi^2 = \sum_{i} \frac{(R(M_i) - \hat R)^2}{ \hat R_0^2}\,,
\end{equation}
with $\hat R$ the universal predictor and $\hat R_0 =12 \mathrm{km}$ a crude estimate of $\hat R$.
Since $R(M) \in [10, 14]\,$km for the majority of astrophysical EoSs, we would find $\chi^2/N_{\dof} \sim(2/10)^2=0.04$.
On the other hand, $\Lambda(M)$ varies over orders of magnitude, and $\Lambda_{1.4} \in [200, 800]$, see Fig.~\ref{fig:R14L14posterior}.  Then the goodness-of-fit
will average to $\chi^2/N_{\dof}\sim (300/500)^2 \sim 0.36$.  
The radius is relatively EoS-independent by this metric under a fractional uncertainty approach; this contrasts with the use of measurement uncertainty as the tolerance factor, where both relations would be comparably poor.  
Therefore, the choice of tolerance factor sensitively impacts what the resulting goodness-of-fit represents.  This is true even when only the fit parameters are of interest, as those will also depend on the tolerance.  

The tolerance factors we use are coarse heuristics for potentially better-motivated choices.  For example, a complete GW simulation study would allow a precise estimate of $\sigma_{\Lambda}$ for a range of binary parameters and detector sensitivities. There are additional choices for the tolerance factor that we do not investigate, such as $\sigma_{\Lambda} = \alpha \delta\Lambda_{stat}^\beta$ , for some (potentially dimensionful) constant $\alpha$ and exponent $\beta$.  Additionally, the tolerance factor may be designed to be agnostic to errors of the fit in certain mass ranges; if for example, sub-solar mass NSs cannot be formed astrophysically, then it is not necessary that the relation is well fit below $M_{\odot}$.  This choice is degenerate, however, with a choice of which NSs the $\chi^2$ is marginalized over; see the discussion in Sec.~\ref{sec:discussion}.

\subsection{EoS set}
\label{sec:EoSset}

The final ingredient of the EoS-independent relation fits is the EoS set and its distribution $P(\epsilon)$. Since our goal is to assess EoS-independence for flexible EoS sets, we use the model-agnostic prior of Ref.~\cite{Essick:2019ldf}, constructed to minimize the impact of nuclear theory input\footnote{\red{Though certain EoS models are used to condition the process, the final EoS distribution depends only weakly on those EoSs which are used for conditioning, see~\cite{Landry:2018prl, Essick:2019ldf}.}}. EoSs are drawn from multiple Gaussian Processes sampling a range of covariance kernels (correlation scale and strength) between different densities.  \red{Each EoS is stitched to a low-density representation of the SLy4 EoS~\cite{Douchin:2001sv} at low densities.} The final EoS \red{prior} predicts NSs with a very wide range of $R\in (8,16)$\,km. We condition this set against radio data~\cite{Antoniadis:2013pzd,Cromartie:2019kug,Fonseca:2021wxt} for the maximum NS mass, and refer to this as the \emph{pulsar-informed set}. We also consider an \emph{astrophysically-informed set}, obtained in~\cite{Legred:2021hdx} by further conditioning on X-ray~\cite{Miller:2019cac,Miller:2021qha,Riley:2019yda,Riley:2021pdl} and GW~\cite{TheLIGOScientific:2017qsa,Abbott:2020uma} data\footnote{\red{The astrophysical data we use are independent of any choice of the EoS and do not use any EoS-independent relations.  Therefore the inclusion of additional data will only improve the quality of fits if the data explicitly favor a set of EoSs which are well fit by the relations.}}.

Due to its flexible construction, both the pulsar-informed and the astrophysically-informed sets contain EoSs with phase transitions, both strong and weak. We therefore further split each set in EoSs without (referred to as the \emph{hadronic set}) and with (referred to as the \emph{mixed-composition set}) phase transitions. In order to identify EoSs with phase transitions, we use the moment-of-inertia-based feature extraction procedure from~\cite{Essick:2023fso}. This procedure can identify both strong and weak phase transitions, including phase transitions that do not result in multiple stable branches or have a large impact on the macroscopic observables.  We set a  high threshold for phase transitions, requiring a change in internal energy per particle of $\Delta(E/N)\geq 30 \mathrm{MeV}$, see Ref.~\cite{Essick:2023fso}.  As before, we also only use EoSs with a single stable branch in the $M$--$R$ relation.  Including EoSs with multiple stable branches would require choices in the construction of the $\chi^2$ to weight each branch and exclude unstable branches, but would likely decrease the goodness-of-fit of the relations to the mixed composition EoS set.  

Finally, for comparison, we repeat the same fits with piecewise-polytropic and spectral EoSs, using the pulsar-informed and astrophysically-informed distributions from Ref.~\cite{Legred:2022pyp}.  \red{We use a 4-parameter piecewise-polytrope parametrization~\cite{Read:2008iy}, with 2 fixed stitching densities, 3 sampled  polytropic indices, and one sampled overall pressure scaling.  For the spectral EoS, we use a 4 parameter EoS (i.e. 4 basis functions in the spectral exponent)~\cite{Lindblom:2010bb}, and the parameter distribution given by Ref.~\cite{Wysocki:2020myz}, which reduces the range of parameter space sampled while significantly improving the fraction of EoS samples which are physically viable.  In both cases, we stitch to a low-parameter representation of the Sly4 EoS, as described in Ref.~\cite{Carney:2018sdv}}.   We follow Ref.~\cite{Carney:2018sdv} in allowing the EoS prior to extend up to $c_s \leq 1.1$, in order to allow an acausal model to represent a potential causal model which is not representable by the parametrization.  In \citet{Legred:2022pyp}, this choice was found to affect the distribution on the piecewise-polytrope EoS; it may additionally affect EoS-independence by allowing additional (unphysical) variation in the EoS.

\section{EoS-independent relations}
\label{sec:results}

We fit a set of proposed EoS-independent relations to different EoS distributions and evaluate their universality.  Throughout, unless otherwise stated, we use a fixed tolerance factor value of $\sigma_\Lambda=210$.\footnote{Simulations suggest that measurement uncertainty in $\Lambda$ is approximately independent of the value of $\Lambda$ (equivalently, the NS mass) and inversely proportional to the signal strength~\cite{Wade:2014vqa}.}  When $\Lambda$ is not predicted by the fit but it is the independent variable of the relation, we propagate the uncertainty through the relation to the dependent quantity.  
For example 
\begin{equation}
    \sigma_I(\Lambda) = \deriv{\tilde I(\Lambda, \bm{\theta}_f)}{\Lambda}\Big|_{\Lambda} \sigma_{\Lambda} \,, 
\end{equation}
where $\bm \theta_f$ are fiducial parameters of the fit, and $\tilde I$ is the EoS-independent predictor of $I$ from $\Lambda$ which depends on $\Lambda$ via the derivative of the predictor.  When neither the independent or dependent variable are $\Lambda$, we use a different strategy; see Secs.~\ref{sec:results-R-Love} and~\ref{sec:results-alphac-c}.  For relations where $\Lambda$ is indeed the dependent quantity and the tolerance factor is constant, this strategy results in optimization problems which are mathematically identical to previous work, e.g.~\cite{Carson:2019rjx}.   Crucially though, now the goodness-of-fit statistic can be interpreted as a measure of EoS-independence relative to observations.  

In this section we show plots for the nonparametric-mixed and spectral astrophysically-informed EoS distributions.  
We display additional plots for the piecewise-polytrope and hadronic-nonparametric EoS distributions as well as fit parameters in Appendix~\ref{app:additional-figures}.

\subsection{I--Love--Q}
\label{sec:results-I--Love--Q}

We begin with the I--Love--Q relations~\cite{Yagi:2013awa} for the dimensionless quadrupole moment $Q$, moment of inertia $I$, and tidal deformability $\Lambda$ of a NS.
The existence of such relations, at least approximately, may not be surprising.  In Newtonian gravity, for example, the quadrupole moment can be computed from the moment of inertia exactly.
In GR, however, the definitions of these quantites do not coincide, which is to say the relationship of angular momentum, angular velocity, and the second multipole of the gravitational field is nontrivial for slowly-spinning compact objects~\cite{Hartle1967}. 

We use a slightly modified form for the I--Love--Q relations compared to Ref.~\cite{Yagi:2013awa}, which was shown by Ref.~\cite{Carson:2019rjx} to produce better behavior in the Newtonian limit:  
\begin{align}
    \label{eq:i-love-relation}
    \hat I(\Lambda; a,b,K_{yx}) &= K_{yx} \Lambda^{\alpha} \frac{1 + \sum_{i=1}^{3} a_i \Lambda^{-i/5}}{1 + \sum_{i=1}^{3} b_i \Lambda^{-i/5}}\,, \\
    \label{eq:q-love-relation}
    \hat Q(\Lambda; a,b,K_{yx})&=   K_{yx} \Lambda^{\alpha} \frac{1 + \sum_{i=1}^{3} a_i \Lambda^{-i/5}}{1 + \sum_{i=1}^{3} b_i \Lambda^{-i/5}}\,,\\
    \label{eq:i-q-relation}
    \hat I(Q; a,b,K_{yx}) &=  K_{yx} Q^{\alpha} \frac{1 + \sum_{i=1}^{3} a_i Q^{-i/5}}{1 + \sum_{i=1}^{3} b_i Q^{-i/5}}\,.
\end{align} 
Here, $a_i, b_i$, and $K_{yx}$ are free parameters which are fit. 
These forms ensure that when $a_i$ and $b_i$ are zero, these relations limit to the Newtonian form. We display best-fit parameters in Table~\ref{tab:I--Love--Q-coefficients-np-sp}.

\begin{table}[]
    \centering
    $\chi^2/N_{\dof}$\\
    \begin{tabular}{|c|c|c|c|}
    
    \hline
     \backslashbox{EoS Dist.}{Relation}& $I(\Lambda)$ &$I(Q)$ & $Q(\Lambda)$\\
     
     \hline
    \nphadrlabel \,(astro) & $\nphadrilpostcost$ & $\nphadriqpostcost$ & $\nphadrqlpostcost$ \\
    \nphadrlabel \,(psr) & $\nphadrilpsrcost$ & $\nphadriqpsrcost$ & $\nphadrqlpsrcost$ \\
    \npmixedlabel \,(astro) & $\npmixilpostcost$ & $\npmixiqpostcost$ & $\npmixqlpostcost$ \\
    \npmixedlabel \,(psr) & $\npmixilpsrcost$ & $\npmixiqpsrcost$ & $\npmixqlpsrcost$ \\
    \splabel \,(astro) & $\spilpostcost$ & $\spiqpostcost$ & $\spqlpostcost$ \\
    \splabel \,(psr) & $\spilpsrcost$ & $\spiqpsrcost$ & $\spqlpsrcost$ \\
    \pplabel \,(astro) & $\ppilpostcost$ & $\ppiqpostcost$ & $\ppqlpostcost$ \\
    \pplabel  \,(psr) & $\ppilpsrcost$ & $\ppiqpsrcost$ & $\ppqlpsrcost$ \\
    \hline
     \end{tabular}
    \caption{Table $\chi^2/N_{\dof}$ for the I--Love--Q relations for several EoS distributions.  Here GP represents the nonparametric (Gaussian Process) distributions, SP represents the spectral distributions, and PP represents the piecewise-polytrope distributions.  We show results for each of the pulsar-informed distributions (psr), and fully astrophysically-informed distributions (astro).}
    \label{tab:I--Love--Q-costs}
\end{table}

\begin{figure*}[htp!]

    \centering
    \includegraphics[height=7.5cm, width=.32\textwidth]{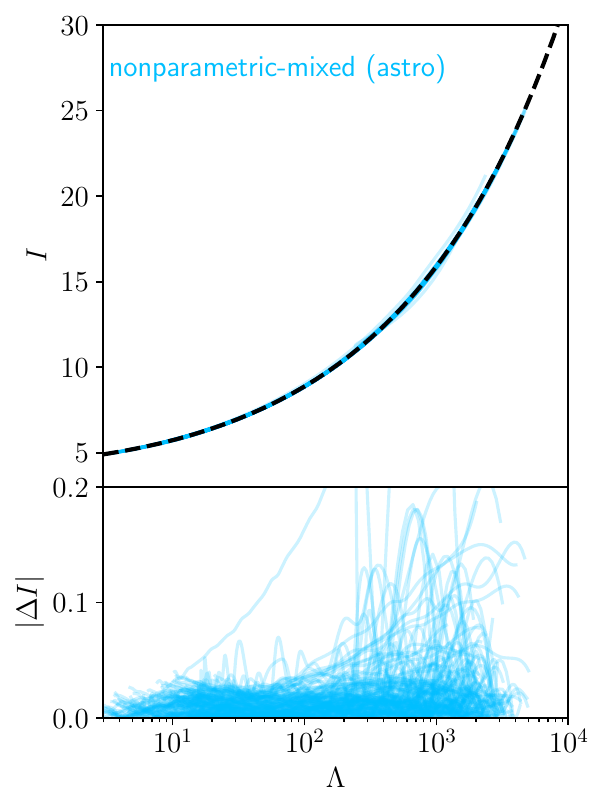}
    \includegraphics[height=7.5cm, width=.32\textwidth]{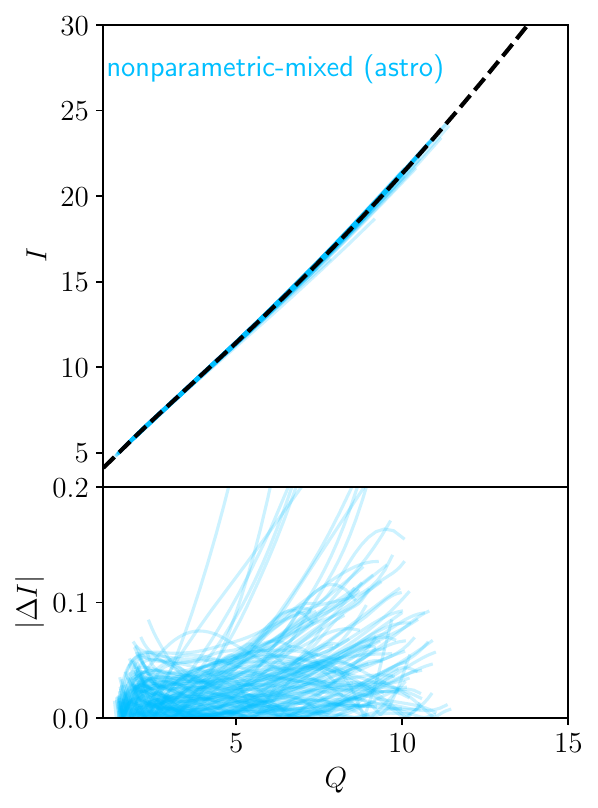}
    \includegraphics[height=7.4cm, width=.32\textwidth]{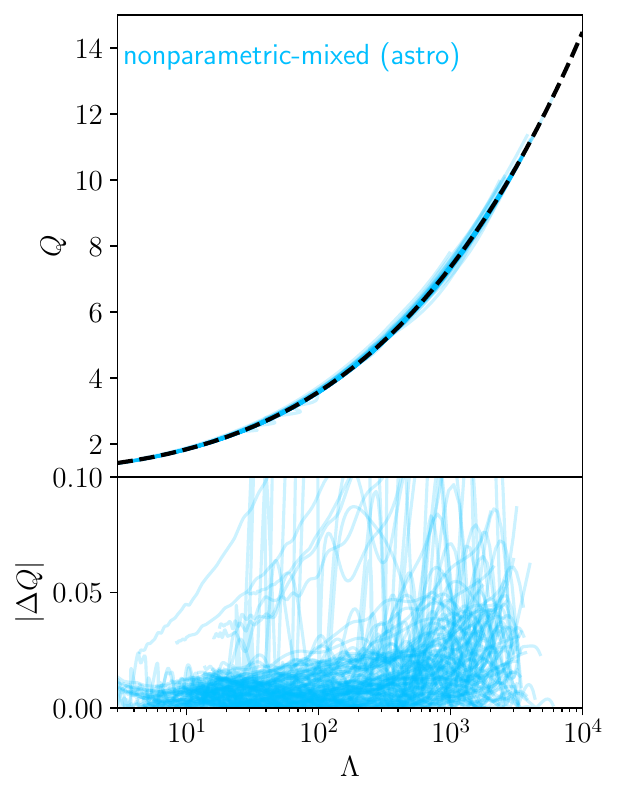}
    
    \caption{\emph{Top, each panel}: EoSs drawn from nonparametric mixed-composition EoS distribution conditioned on all astrophysical data (in blue), along with the best-fits (black dashed).  From left to right we display the I-Love, I-Q, and Q-Love relations. 
    \emph{Bottom, each panel}: Residuals of the fit relative to each of the sampled EoSs.  This represents a measure of the ``error" of using the particular relation with the given EoS set.}
 
    \label{fig:I--Love--Q-fits-np-mixed-post}
\end{figure*}

\begin{figure*}[htp!]
    \centering
    \includegraphics[height=7.5cm, width=.32\textwidth]{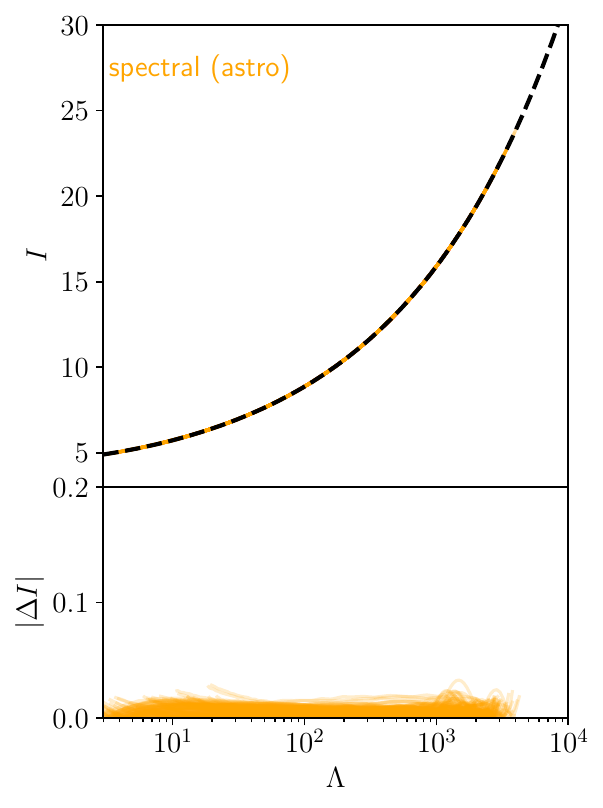}
    \includegraphics[height=7.5cm, width=.32\textwidth]{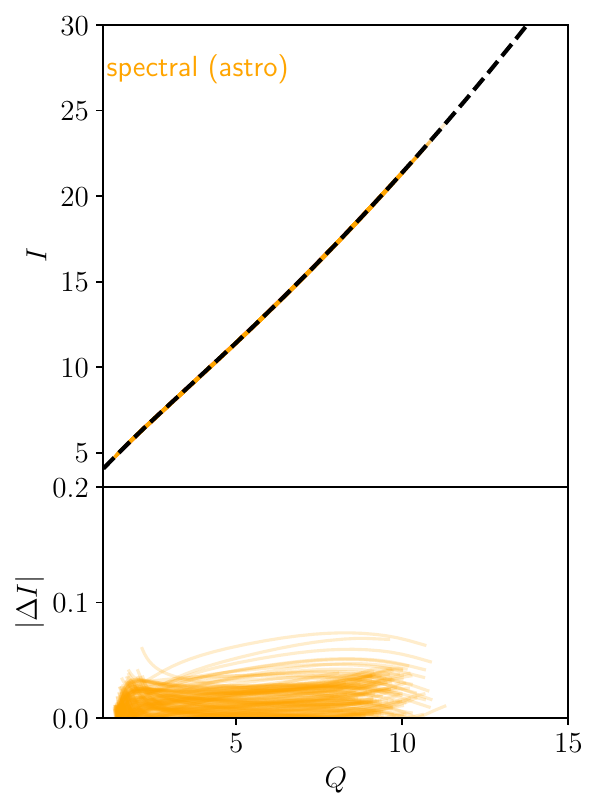}
    \includegraphics[height=7.4cm, width=.32\textwidth]{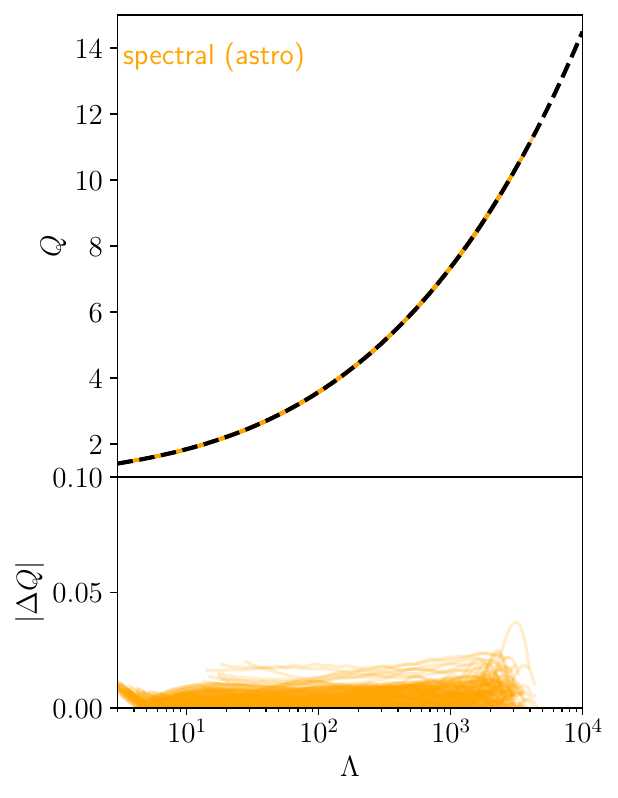}
    
    \centering
    \caption{The same as Fig.~\ref{fig:I--Love--Q-fits-np-mixed-post} but with the spectral EoS distribution conditioned on all astrophysical data. We use identical axes ranges between the two figures. Worst-case residuals are of order 10 times smaller than the nonparametric mixed-composition distribution seen in Fig.~\ref{fig:I--Love--Q-fits-np-mixed-post}.
    }
    \label{fig:I--Love--Q-fits-sp-post}
\end{figure*}

We solve the TOV equations in the slow-rotation limit up to second order~\cite{Hartle1967} to compute the dimensionless moment of inertia, quadrupole moment, and tidal deformability\footnote{We thank Victor Guedes for the use of code to solve the TOV equations in the slow-rotation limit up to second order.}.  We then fit the parameters  of each relation using a nonlinear least squares algorithm.  
 We display the \emph{loss}, i.e., best fit $\chi^2/N_{\rm dof}$ value,  of each fit for each EoS distribution in Table~\ref{tab:I--Love--Q-costs}.  In this context $N_{\rm dof}$ represents the number of degrees of freedom in the data, which is the number of points fit (200) minus the number of fitted parameters.  
 The loss measures the residuals in the fit relative to $\sigma_{\Lambda} = 210$, as described in Sec.~\ref{sec:tolerance}.   
 
 The I--Love--Q relations hold independent of EoS distribution to very high precision, with loss values less than $3\e{-3}$ for almost all relations.   In particular \red{$I(Q)$}, with losses of $\lesssim 10^{-5}$ indicates that even with ${\cal{O}}(10)$ improvement in GW detector sensitivity, the systematic error of the relation will still be at sub-percent level compared to statistical uncertainties.  Nonetheless,  the parametric EoS distributions display moderately better EoS-independence  than the corresponding nonparametric distributions, typically by a factor of 3-10.   Similarly,   the hadronic nonparametric distribution is typically a factor of 2-3 better than the corresponding mixed-composition distributions.  In all cases, the fits to pulsar-informed distributions show higher losses than the ones conditioned on all astrophysical data.  For the spectral distributions, the difference is marginal, about a factor of 2, whereas for the nonparametric mixed distribution the difference is almost a factor of 20 for the relations involving $\Lambda$.

We display the fits for the nonparametric mixed-composition, and spectral EoS distributions in Figs.~\ref{fig:I--Love--Q-fits-np-mixed-post} and~\ref{fig:I--Love--Q-fits-sp-post} respectively, again each distribution conditioned on all astrophysical data.
For other distributions, see Appendix~\ref{app:additional-figures}.  The higher degree of EoS-independence in the spectral fit is apparent in the residuals, which are several times smaller than the nonparametric residuals.

The $I$--$Q$ relation shows the smallest loss in EoS-independence (by a factor of 10) when moving from the spectral pulsar-informed distribution to the equivalent nonparametric distribution.  This indicates that the $I$--$Q$ relation is fundamentally more EoS-independent than relations involving $\Lambda$. This is potentially related to the discussion of emergent symmetries in Ref.~\cite{Yagi:2014qua}, which demonstrated that the $I$--$Q$ relation is indeed EoS-independent  under the elliptical isodensity approximation, which is nearly true in astrophysically relevant NSs~\cite{Yagi:2014qua}.

\subsection{Binary-Love}
\label{sec:binary-love}

\begin{figure}
    \centering
    \includegraphics[width=.49\textwidth]{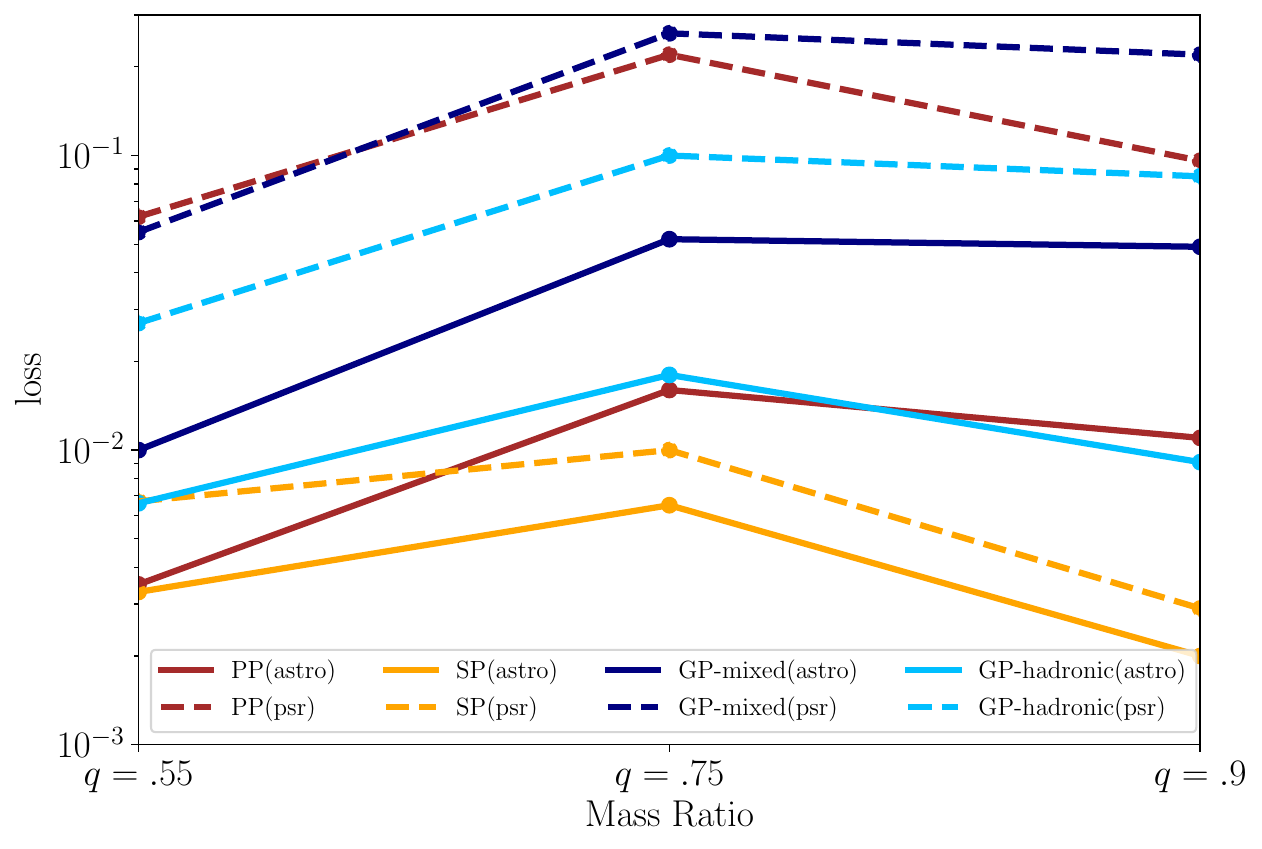}
    \caption{The costs shown in Table~\ref{tab:binary-love-costs}.  The spectral costs are in general the lowest, especially for more equal-mass binaries.
    }
    \label{fig:binary-love-costs}
\end{figure}

The \emph{Binary-Love} relation  allows us to estimate the tidal deformability of one NS in a binary given its NS companion's deformability.  The expression is given in terms of the symmetric deformability $\Lambda_s \equiv (\Lambda_1 + \Lambda_2)/2$ and the antisymmetric deformability, $\Lambda_a = (\Lambda_2 - \Lambda_1)/2$ where $\Lambda_1$ and $\Lambda_2$ are the deformabilities of two NSs~\cite{Yagi:2015pkc}:
\begin{equation}
    \label{eq:binary-love-relation}
    \Lambda_a(\Lambda_s, q; b, c) = F_n(q)\Lambda_s\frac{1+\sum_{i=1}^3\sum_{j=1}^2 q^jb_{ij}\Lambda_s^{-1/5}}{1+ \sum_{i=1}^3 \sum_{j=1}^2q^jc_{ij}\Lambda_s^{-1/5}}\,.
\end{equation}
Here $b_{ij}$, and $c_{ij}$ are parameters which are fit.  
For the Binary-Love relation, we use a NS distribution truncated at $0.8 M_{\odot}$ rather than $1.0M_{\odot}$, this is necessary to allow the relationship to be evaluated over a wider range of mass ratios $q$. 
Fit coefficients for the astrophysically-informed EoS sets are given in Table~\ref{tab:binary-love-coefficients-np-sp}. Here, and in the rest of the paper, we solve the TOV equations only to first order in the small spin parameter using the approach from \cite{Landry:2014jka}.
We display the fit losses in Table~\ref{tab:binary-love-costs}, and additionally plot them in Fig.~\ref{fig:binary-love-costs}. 

The losses are noticeably higher than any of the I--Love--Q relations for corresponding EoS sets, indicating the relation is less EoS-independent. The use of lower mass cutoff inevitably leads to an increase in loss for the relation, as low-mass NSs have larger tidal deformabilities; however, raising the mass cutoff to $M_{\min} = 0.9$ lowers losses by only a factor of $\sim 2-3$.   This indicates that the fits are indeed worse than the I--Love--Q relations.  

\begin{figure*}
    \centering
    \includegraphics[width=.49\textwidth]{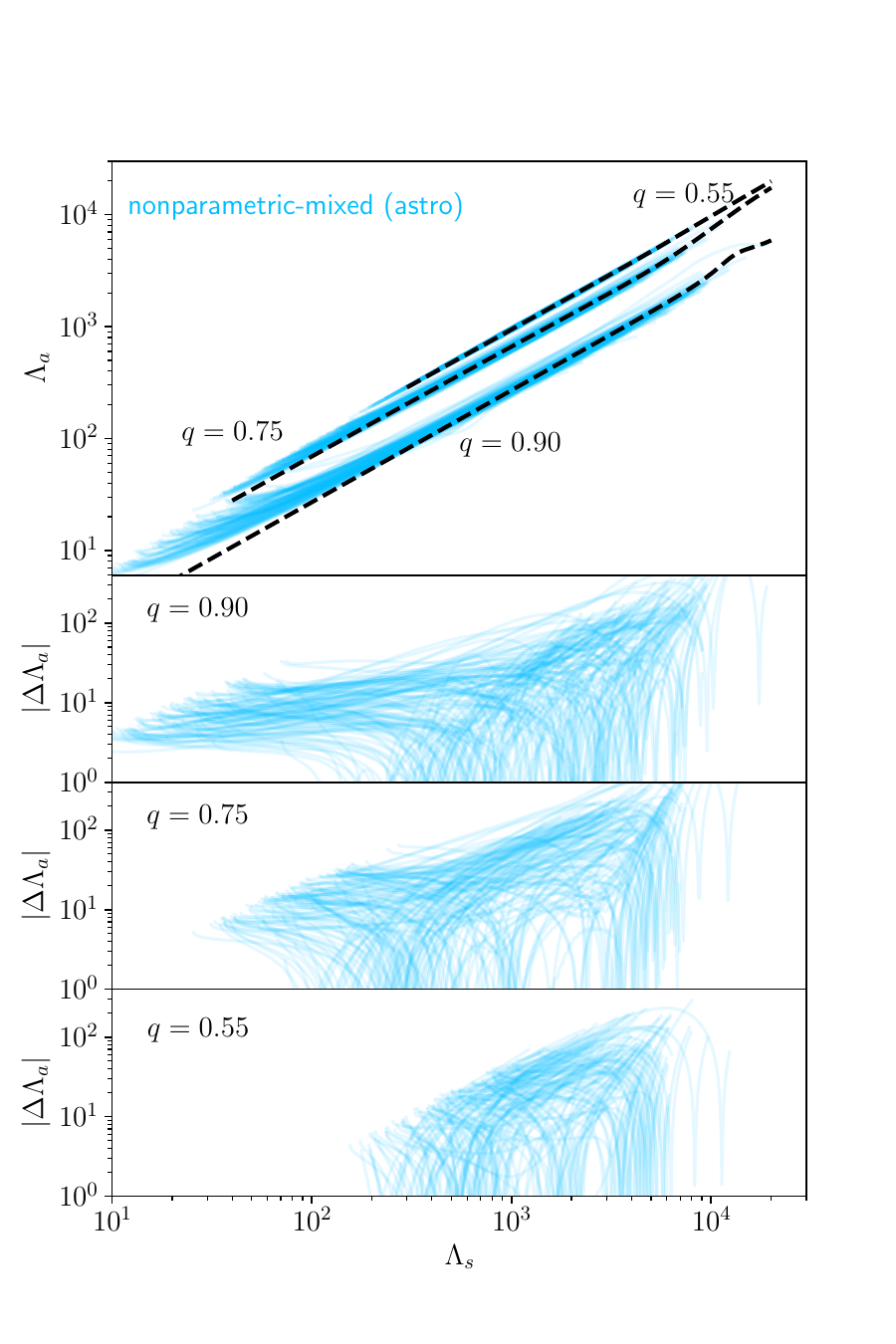}
    \includegraphics[width=.49\textwidth]{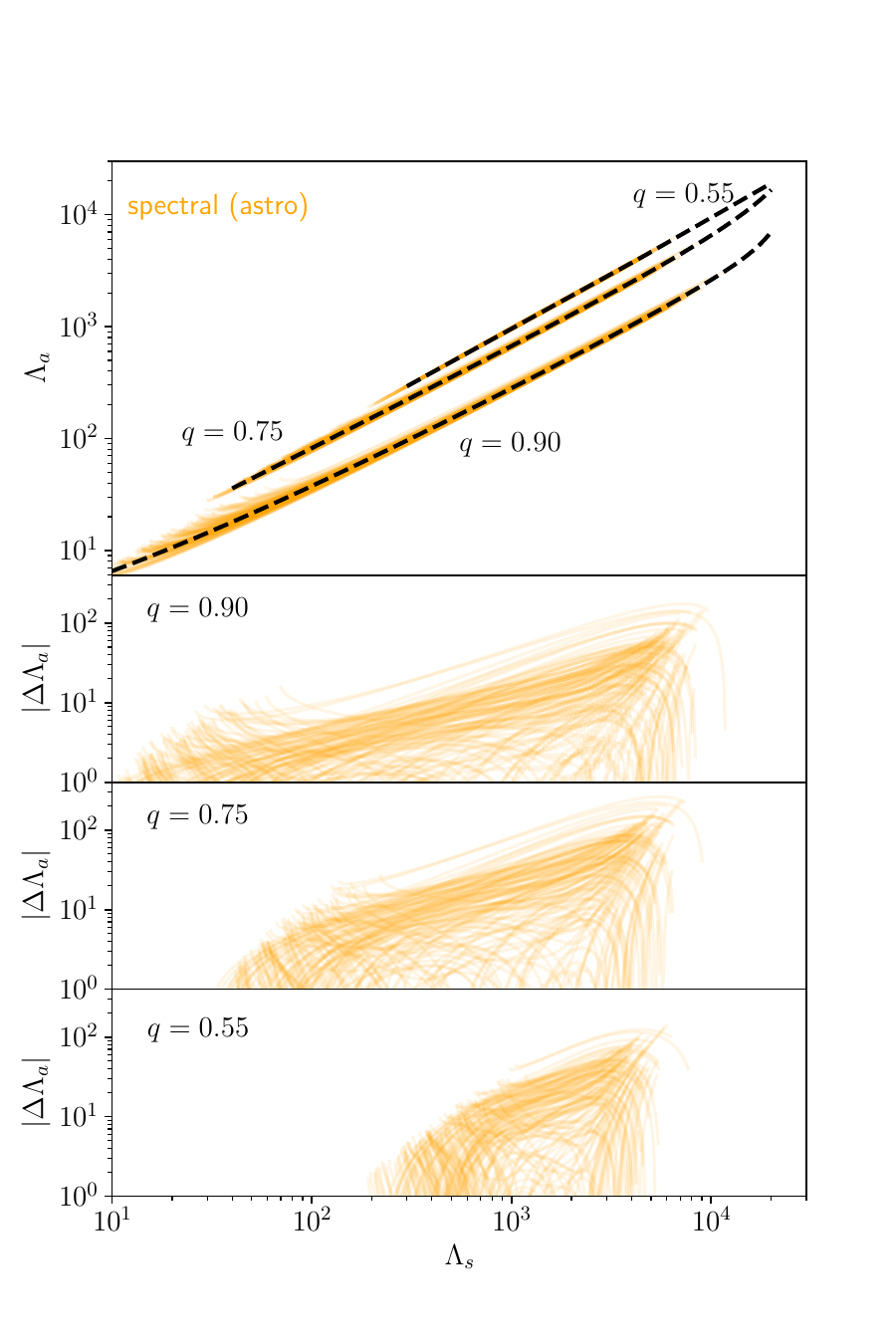}
    \caption{Similar to Fig.~\ref{fig:I--Love--Q-fits-np-mixed-post}. \emph{Left}: The Binary-Love relation fitted, along with all EoSs for the nonparametric EoS distribution with mixed composition when conditioning on all astrophysical data.  We plot each fit for three different mass ratios, $q=0.55, q=0.75$, and $q=0.9$. 
    \emph{Right}:  The same for the spectral EoS distribution.}
    \label{fig:binary-love-relations}
\end{figure*}

Nonetheless, the spectral EoS distribution fits are $\sim$50 times better than the nonparametric and piecewise-polytrope distributions.
The better fit to the spectral distribution  might be due to large correlations between density scales.  Such correlations may reduce the variation in $\Lambda$ across mass scales, making the relation between $\Lambda(m_1)$ and $\Lambda(m_2)$ more EoS-independent.   
  The astrophysically-informed fits show improvement over pulsar-informed fits, with typical loss values 3-5 times better.
  The piecewise polytrope is by far the most improved distribution upon inclusion of more data, with losses decreasing by factors of more than 10.  In all cases the hadronic distributions give improved fits relative to the mixed composition distributions, typically by a factor of 10 in loss.  For the worst-fit case, the nonparametric pulsar-informed mixed distribution, the fit quality ($\chi^2/N_{\dof}= \npmixblmedpsrcost$) may be poor enough to pose challenges for current-generation GW detectors, as it indicates systematic errors of order $60\%$ in the predicted value of $\Lambda_a$ relative to statistical uncertainties.  Figure~\ref{fig:binary-love-relations} shows nonparametric mixed and spectral fits relative to sampled EoSs.  The larger variation of the nonparametric EoS set relative to the spectral set is apparent.

\begin{table}[]
    \centering
    $\chi^2/N_{\dof}$\\
    \begin{tabular}{|c|c|c|c|}
    
    \hline
     \backslashbox{EoS Dist.}{Relation}& $q=0.55$ &$q=0.75$&$q=0.9$\\
     
     \hline
    \nphadrlabel \,(astro) & $\nphadrbllowpostcost$&$\nphadrblmedpostcost$&$\nphadrblhipostcost$\\
    \nphadrlabel \,(psr) & $\nphadrbllowpsrcost$&$\nphadrblmedpsrcost$&$\nphadrblhipsrcost$\\
    \npmixedlabel \,(astro) & $\npmixbllowpostcost$&$\npmixblmedpostcost$&$\npmixblhipostcost$\\
    \npmixedlabel \,(psr) & $\npmixbllowpsrcost$&$\npmixblmedpsrcost$&$\npmixblhipsrcost$\\
    \splabel \,(astro) & $\spbllowpostcost$&$\spblmedpostcost$&$\spblhipostcost$\\
    \splabel \,(psr) & $\spbllowpsrcost$&$\spblmedpsrcost$&$\spblhipsrcost$\\
    \pplabel \,(astro) & $\ppbllowpostcost$&$\ppblmedpostcost$&$\ppblhipostcost$\\
    \pplabel \,(psr) & $\ppbllowpsrcost$&$\ppblmedpsrcost$&$\ppblhipsrcost$\\
    \hline
     \end{tabular}
    \caption{Table $\chi^2/N_{\dof}$ for the Binary-Love relations for several distributions on the EoS and binary mass ratios.}
    \label{tab:binary-love-costs}
\end{table}

The large differences in fit quality for nonparametric distribution with mixed composition and hadronic composition are consistent with observations of the Binary-Love relation (as presented here) failing to describe effectively EoSs with phase transitions \cite{Carson:2019rjx, Tan:2021nat}.  This could potentially lead to analyses using Binary-Love relations artificially downranking EoSs which support hybrid stars.  This effect will likely be smaller than an e-fold in likelihood for any individual event, but such effects may multiply in a hierarchical analysis, leading to large errors after many events.

\subsubsection{Changing fit quality with the tolerance factor}

\label{sec:binary-love-tolerance-change}

\begin{figure}
    \centering
    \includegraphics[width=.49\textwidth]{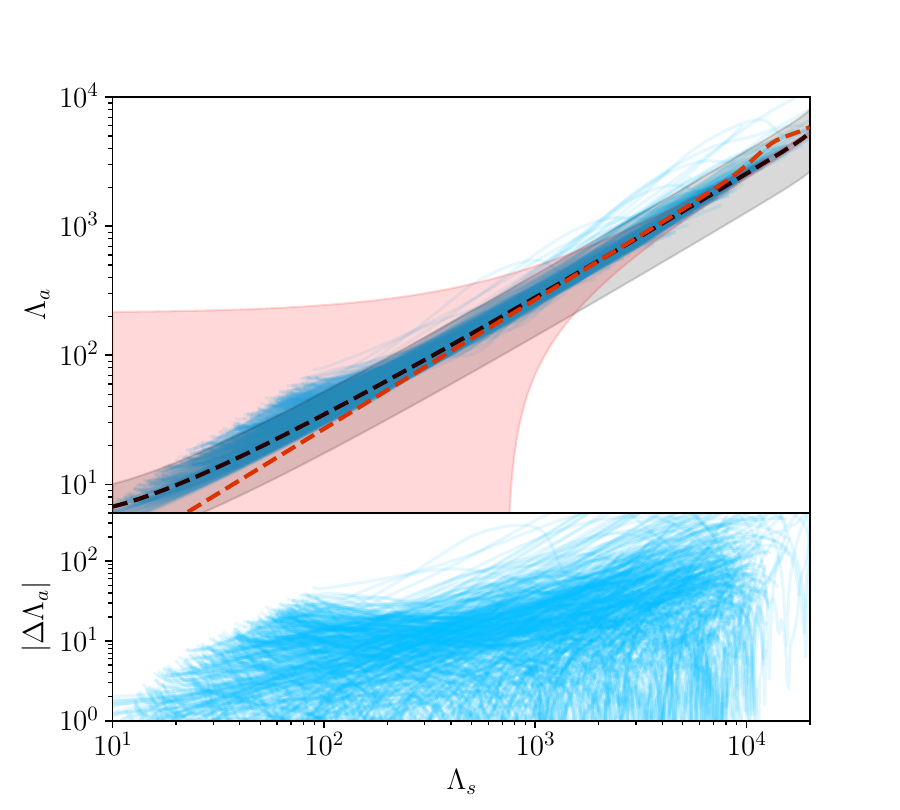}
    \caption{The Binary-Love fit to the mixed nonparametric, astrophysically-informed EoS distribution when applying a modified tolerance factor that favors better fits at low-$\Lambda$ values.  The best fit line is in dashed black, plotted over draws from the nonparametric distribution in blue.  For comparison, we plot in dashed red the best-fit line for the uniform tolerance factor fit to the same distribution, the same as Fig.~\ref{fig:binary-love-relations}.   We shade the $\sigma(\Lambda_s)/2$ area away from best fit $q=0.9$ curve in pink for the uniform tolerance factor, and in gray for the modified, constant relative tolerance factor.  The fit requires better agreement at low $\Lambda_a$ in order to achieve low cost, and therefore tit appears better by eye than the fit in Fig.~\ref{fig:binary-love-relations}, especially on a log-log plot.}
    \label{fig:binary-love-tolerance-change}
\end{figure}

The deteriorating quality of the fit at low values of $\Lambda_s$ is apparent in Fig.~\ref{fig:binary-love-relations}, left panel.
 This is because assuming a constant tolerance factor for $\Lambda$ upweights relative errors where $\Lambda$ is large, i.e. the regime where small relative differences lead to very large $\chi^2 $ values.  
 It is possible to use the tolerance factor to improve the fit quality at low $\Lambda$.  Instead of choosing an observational value for the tolerance factor $\sigma(\Lambda_s)$, we set $\sigma(\Lambda_s)= \Lambda_a(\Lambda_s)$.  Such a fit may be useful for tidal analyses of binaries containing a massive NS, as it gives constant relative uncertainty and therefore tolerates only small errors in $\Lambda_a$ when $\Lambda_a$ itself is small. We plot the fit achieved in Fig.~\ref{fig:binary-love-tolerance-change}, scaling the uncertainty by a factor of $0.5$ for display purposes.  We additionally plot the region encompassed by $\pm \sigma(\Lambda_s)$ and shade the region in between for the $q=0.9$ fit.  This demonstrates the role of tolerance factors and the flexibility they offer.

\subsection{C--Love}
\label{sec:results-c-lambda}

Another established EoS-independent relation relates compactness to tidal deformability~\cite{Maselli:2013mva, Yagi:2013awa}.  This relation is useful for determining the radii of NS with measured tidal deformabilities and masses from GW observations.  
Such a relation is plausible: radius and tidal deformability are linked by definition
\begin{equation}
    \label{eq:lambda-def}
    \Lambda = \frac{2}{3}k_2(m)\quant{R/m}^5 = \frac{2}{3} k_2 C^{-5}\,,
\end{equation}
though a truly EoS-independent description would require $k_2$, the tidal Love number, to be either  independent of the EoS or expressible only as a function of $C$.

The relation  is given as follows, again using a fitting form from Ref.~\cite{Carson:2019rjx} 
\begin{equation}
    \label{eq:c-lambda-relation}
    C = K_{C\Lambda} \frac{1 + \sum_{i=1}^3a_i\Lambda^{-i/5} } {1 + \sum_{i=1}^3b_i\Lambda^{-i/5}}\,.
\end{equation}
Similar to the I-Love-Q case, $a_i, b_i$, and $K_{C\Lambda}$ are parameters to be fit.  
As before, we propagate a constant $\Lambda$ uncertainty to a  $C$ uncertainty.  However, for high-compactness stars, GWs are expected to be weakly informative probe, leading to poor fits for the high-compactness part of the relation.  Moreover, X-ray probes of compactness can provide complementary constraints~\cite{Riley:2019yda,  Miller:2019cac}.  We therefore hybridize two tolerance factors:  
\begin{equation}
    \sigma_C^{-2} = \sigma_{C, \rm {x-ray}}^{-2} + \sigma_{C, \rm {GW}}^{-2}\,.
\end{equation}
The X-ray uncertainty $\sigma_{C, \rm {x-ray}}^{-2}$ is negligible for $C\lesssim 0.16$, while the GW uncertainty $\sigma_{C, \rm{GW}}$ is negligible for $C\gtrsim 0.2$.  This corresponds to a transition from X-ray to GW data dominating constraints near $\Lambda=200-500$.  The total tolerance factor is not representative of any particular measurement, but rather provides a holistic picture of the statistical uncertainty.    

Results are shown in Table~\ref{tab:C--Love-costs}.
We find the fit qualities to be $\sim 100$ times  poorer than for the I--Love--Q relations, even for the parametric distributions.  Contrary to the Binary-Love case, the C--Love goodness-of-fit is relatively independent of conditioning on additional data, with the loss changing by $\lesssim 2$ in all cases when additional astrophysical data are included. Also, for the nonparametric EoSs, the C--Love relation is not appreciably better fit to the hadronic distribution than to the mixed distribution.    Similarly to the Binary-Love  relations, the mixed-composition nonparametric distribution conditioned only on heavy pulsar mass measurements shows a loss of $3.6\e{-1}$, indicating systematic errors are already comparable to statistical uncertainties.  The same holds true for the piecewise-polytrope distribution, though the piecewise-polytrope loss decreases by almost a factor of 10 upon the introduction of additional astrophysical data, while the nonparametric distribution decreases by only a factor of 3.  This is consistent with the discussion in Sec.~\ref{sec:binary-love}, and indicates that the large variance in the piecewise-polytrope distribution that leads to large losses is not consistent with current astrophsyical data from x-ray pulsars and gravitational waves.

\begin{figure*}
    \centering
    \includegraphics[width=.49\textwidth]{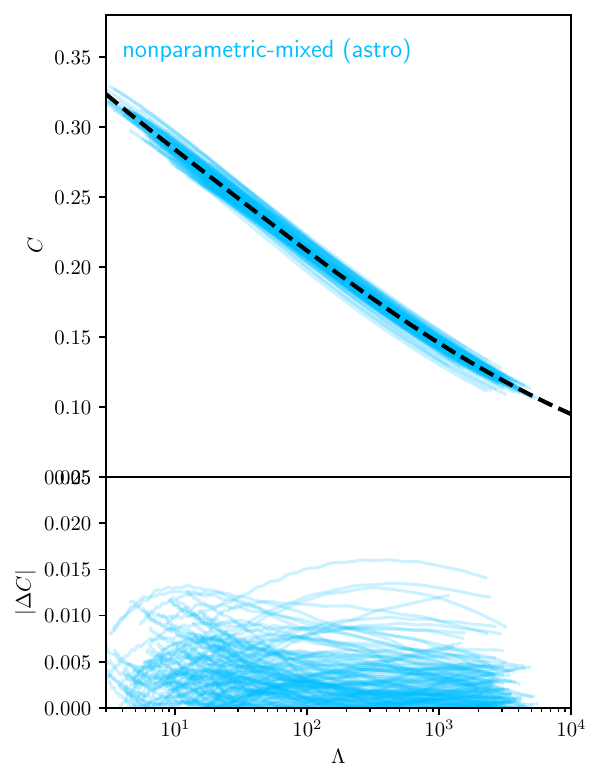}
    \includegraphics[width=.49\textwidth]{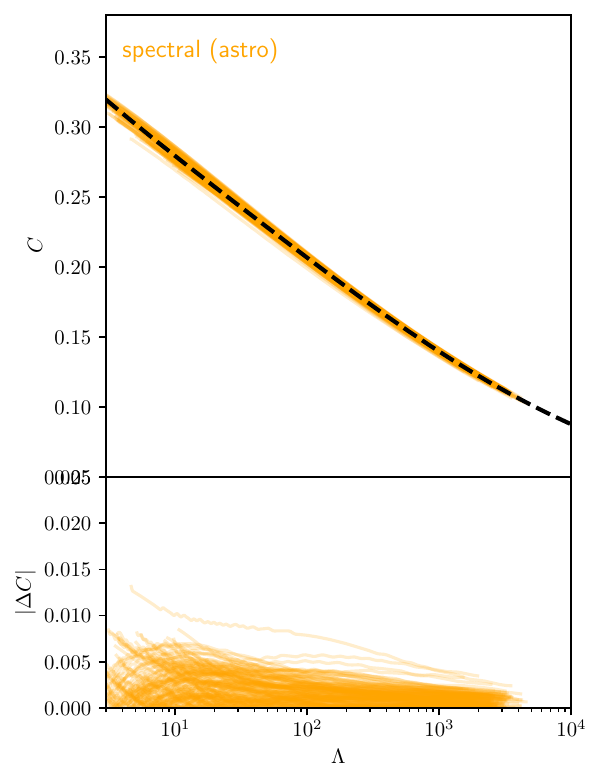}
    \caption{\emph{Left:} The C--Love relation fitted along sampled EoSs for the nonparametric EoS distribution with mixed composition when conditioning on all astrophysical data.
    \emph{Right:}  The same for the spectral parametrization.
    Relative errors are larger for the C--Love relation than for the I--Love--Q relation, and the nonparametric mixed distribution shows greater variability than the spectral distribution.}
    \label{fig:C-Love-relations}
\end{figure*}

Finally, we also display the fits to the nonparametric mixed-composition and spectral EoS distributions conditioned on all astrophysical data in Fig.~\ref{fig:C-Love-relations}.  The nonparametric EoSs have  residuals larger by about a factor of 2, as in the previous examples. 
Fit parameters are given in table~\ref{tab:C-Love-coefficients-np-sp}.

\begin{table}
    \centering
    $\chi^2/N_{\dof}$\\
    \begin{tabular}{|c|c|}
    
    \hline
     \backslashbox{EoS Dist.}{Relation}& $C(\Lambda)$\\
     
     \hline
    \nphadrlabel \,(astro) & $\nphadrclpostcost$\\
    \nphadrlabel \,(psr) & $\nphadrclpsrcost$\\
    \npmixedlabel \,(astro) & $\npmixclpostcost$\\
    \npmixedlabel \,(psr) & $\npmixclpsrcost$\\
    \splabel \,(astro) & $\spclpostcost$\\
    \splabel \,(psr) & $\spclpsrcost$\\
    \pplabel \,(astro) & $\ppclpostcost$\\
    \pplabel \,(psr) & $\ppclpsrcost$\\
    
    \hline
    
     \end{tabular}
    \caption{Table $\chi^2/N_{\dof}$ for the C--Love relations for several distributions on the EoS.}
    \label{tab:C--Love-costs}
\end{table}

The relatively large losses in the $C$--Love relation are consistent with the existence of doppelgangers~\cite{Raithel:2022efm,Raithel:2022aee}: EoSs with similar $\Lambda$ across the parameter space, $\Delta \Lambda <30$, but different $R$, $\Delta R$ up to $0.5$\,km. This phenomenon is due to variability in the EoS at densities below $2\rho_{\rm nuc}$;  the nonparametric EoS prior contains a wide range of low-density behaviors and thus produces EoSs with similar features.    Approximating the nonparametric EoS distribution with this relation may result in errors in compactness $\Delta C \sim 0.02$ , although typical errors are $\Delta C \lesssim 0.01$.  Choosing a fiducial NS radius of 10.5\,km, and a fiducial mass of 1.4\,$M_\odot$, this error can be translated to maximal radius uncertainty of $\sim  1\mathrm{km}$, with typical errors half that, in line with Refs.~\cite{Raithel:2022efm,Raithel:2022aee}.   The presence of these features is additionally consistent with  Fig.~\ref{fig:R14L14posterior}; the nonparametric EoS distribution shows a less EoS-independent relations between $R$ and $\Lambda$. This  indicates that independent radius and tidal deformability measurements will be required in order to effectively constrain the EoS at intermediate ($\sim 1-2\rho_{\rm nuc}$) densities.  

\subsection{$R_{1.4}$--$\tilde \Lambda$}
\label{sec:results-R-Love}

An additional relation between NS tidal properties and the radius  has been proposed by Refs.~\cite{De:2018uhw, Raithel:2018ncd, Zhao:2018nyf}.  The relation leverages the relative insensitivity of $\tilde \Lambda$, the leading order tidal parameter in the post-Newtonian expansion of the GW phase~\cite{Flanagan:2007ix, Wade:2014vqa} as a function of $q$, and the relation given in Eq.~\eqref{eq:lambda-def}: $\tilde \Lambda(R_{1.4}, \mathcal{M}_c)$.  That is, we write the tidal deformability as a EoS-independent function of typical star radius and the chirp mass, $\mathcal{M}_c \equiv (m_1 m_2)^{3/5} / (m_1 + m_2)^{1/5}$, of the binary. 

The relation is given, following Ref.~\cite{Zhao:2018nyf}, by
\begin{equation}
    \label{eq:r1p4-lambda-tilde}
    \frac{a}{\tilde \Lambda} \quant{\frac{R_{1.4}}{\mathcal{M}_c}}^6 = 1\,. 
\end{equation}
 For it to be useful, it should hold for some (perhaps narrow) range of mass ratios, chirp masses, and for a wide range of EoSs for some constant $a$.    In practice, the relation is used to infer $R_{1.4}$ so we use this to define the uncertainty in this case (unlike all other examples), we can no longer only propagate uncertainty from $\Lambda$ measurements because the  chirp mass is also uncertain.  We assume a fiducial uncertainty of $\Delta R_{1.4} = 1.0$\,km, which represents a $\sim \pm 8\%$ measurement of the radius of a NS and use a typical $R_{1.4}$ value of $12$\,km; we select a fiducial grid of $\mathcal{M}_c$ for each EoS, induced by requiring both components to be below $M_{\max}$ and above $1.05M_{\odot}$.  We find that additionally by fixing the chirp mass, the loss of the fit may decrease by a factor of 5 for the spectral distribution, but by only a factor of 2 for the nonparametric distribution, and a factor of 1.5 for the piecewise-polytropic distribution.    The $\chi^2$ in this case is then
\begin{equation}
    \chi^2(a) = \sum_i \sum_j P(\epsilon_i)\frac{ \quant{\frac{1}{a}\tilde\Lambda^{1/6}_{(i)} \mathcal{M}^{(j)}_c - R^{(i)}_{1.4}}^2}{\Delta R_{1.4}^{2}}\,.
\end{equation}
Where  $R_{1.4}^{(i)}$  depends on the EoS($\epsilon_i$), and $\tilde \Lambda_{(i,j)}$ depends on the EoS and the binary parameters, but $R_{1.4}$, the fiducial radius value, is independent of both EoS and binary parameters.  
In this case the optimal solution can be obtained analytically by differentiating the cost with respect to $1/a$.
The loss is then given by $\chi^2(a_*)$, \red{with $a_*$ the optimal solution;} it shown in Table~\ref{tab:r-lambdat-costs}. 
Fit parameters are given in Table~\ref{tab:rtyp-lambdatilde-coefficients-np-sp}.

\begin{table}[ht!]
    \centering
    $\chi^2/N_{\dof}$\\
    \begin{tabular}{|c|c|}
    
    \hline
     \backslashbox{EoS Dist.}{Relation}& $R_{1.4}$--$\tilde \Lambda$\\
     
    \hline
    \nphadrlabel \,(astro) & $\nphadrrlambdatpostcost$ \\
    \nphadrlabel \,(psr) & $\nphadrrlambdatpsrcost$ \\
    \npmixedlabel \,(astro) & $\npmixrlambdatpostcost$ \\
    \npmixedlabel \,(psr) & $\npmixrlambdatpsrcost$ \\
    \splabel \,(astro) & $\sprlambdatpostcost$ \\
    \splabel \,(psr) & $\sprlambdatpsrcost$ \\
    \pplabel \,(astro) & $\pprlambdatpostcost$ \\
    \pplabel \,(psr) & $\pprlambdatpsrcost$ \\
    
    \hline
    
     \end{tabular}
    \caption{Table $\chi^2/N_{\dof}$ for the $R_{1.4}-\tilde \Lambda$  relations for several distributions on the EoS.}
    \label{tab:r-lambdat-costs}
\end{table}

The spectral EoS distributions again show greater levels of EoS-independence than the nonparametric distribution,  indicating a tighter relationship between $R_{1.4}$ and $\Lambda$ in the spectral model,  consistent with Fig.~\ref{fig:R14L14posterior}.  However, fits are typically poorer relative to the I--Love--Q relations and more consistent with the Binary-Love relations.  Similar to the Binary-Love case, the nonparametric and piecewise-polytrope, pulsar-informed distributions show nearly identical loss, $\sim 1.3-1.7\e{-1}$.  The mixed composition distribution shows marginally worse fits, with losses about 1.3 times worse for the pulsar-informed distributions.  When conditioning on additional astrophysical data, the piecewise-polytrope distribution is better fit with the relation, improving by a factor of 2, while the nonparametric distributions improve by less than $25\%$.  This distinction is likely due to relatively strong correlations between $\sim \rho_{\rm nuc}$ and higher densities in the piecewise-polytrope distribution, which are absent in the nonparametric EoS distribution.
See, e.g., Fig. 5 of~\cite{Legred:2022pyp}. 
 These correlations cause  astrophysical measurements to be highly informative at and below nuclear densities in the piecewise-polytrope case, and therefore likely rule out many of the configurations which lead to ``doppleganger"-like behavior~\cite{Raithel:2022efm,Raithel:2022aee}.  This leads to less variation in the relation between $R$ and $\Lambda$ and therefore improves the quality of the fit.  By contrast, there is still a range of low-density behavior within the nonparametric posterior~\cite{Legred:2022pyp}, which likely increases the range of behaviors seen in the $\Lambda-R$ relations of nonparametric EoSs.  This variability would be associated with a lower degree of EoS-independence in the $R_{1.4}$--$\tilde \Lambda$ relation.  

\subsection{$\alpha_c$--$C$}
\label{sec:results-alphac-c}

\begin{figure*}
    \centering
    \includegraphics[width=.48\textwidth]{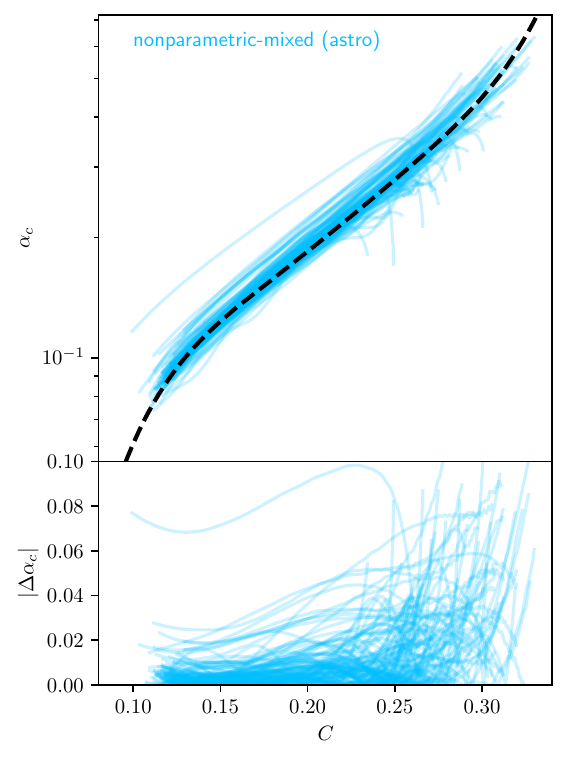}
    \includegraphics[width=.48\textwidth]{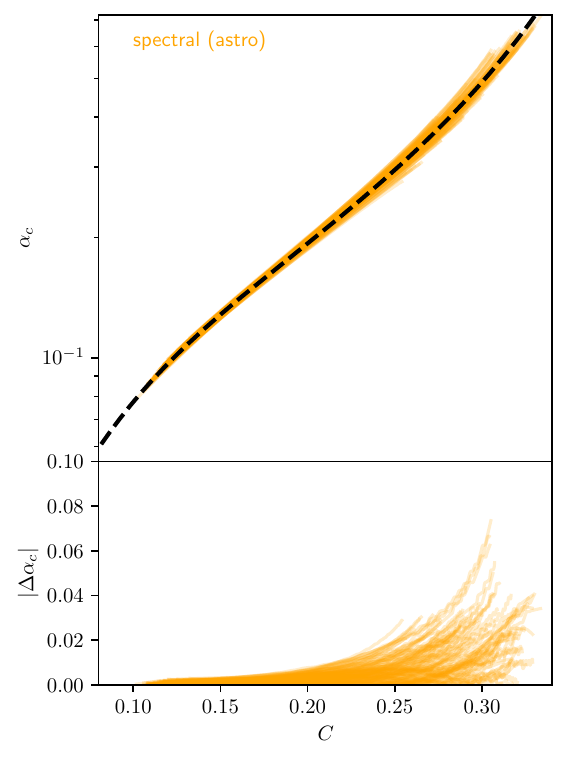}
    \caption{Similar to Fig.~\ref{fig:I--Love--Q-fits-np-mixed-post} but for the $\alpha_c$--$C$ relation.  Left: the relations between $\alpha_c$ and $C$ for the nonparametric EoS model with mixed composition  conditioned on all astrophysical data.
     Right: The same for the spectral parametrization, conditioned on all astrophysical data. }
    \label{fig:alphac-c-relations}
\end{figure*}

A EoS-independent relation between $\alpha_c \equiv p_c/\epsilon_c$ and compactness $C$ was proposed by Ref.~\cite{Saes:2021fzr}.  The quantity $\alpha_c$ is most sensitive  to the EoS only at the highest densities in a star, while the compactness depends on the all densities in the star.
Therefore we would expect EoS parametrizations which impose strong inter-density correlations to be most consistent with the relation.
The expression to be fit is~\cite{Saes:2021fzr}  
\begin{equation}
    \label{eq:c-alphac-relation}
    \ln(\alpha_c) = \sum_{j=0}^5 a_j \ln(C)^5\,.
\end{equation}
The parameters to be fit are $a_j$.
We define a tolerance factor for this relation by propagating the uncertainty in $\Lambda$ through the C--Love relation, and then through the $\alpha_c$--$C$ relation, using fiducial parameters for the C--Love relation given by~\cite{Carson:2019rjx} and for the $\alpha_c$--$C$ relation given by~\cite{Saes:2021fzr}.
In Fig.~\ref{fig:alphac-c-relations} we display the fit and residuals of this relation to our nonparametric, mixed composition, astrophysically-informed EoS distribution, and to the corresponding astrophysically-informed spectral EoS distribution.
Fit coefficients are given in Table~\ref{tab:c-alphac-coefficients-np-sp}.

\begin{table}[ht!]
    \centering
    $\chi^2/N_{\dof}$\\
    \begin{tabular}{|c|c|}
    
    \hline
     \backslashbox{EoS Dist.}{Relation}& $\alpha_c(C)$\\
     
    \hline
    \nphadrlabel \,(astro) & $\nphadralphaccpostcost$ \\
    \nphadrlabel \,(psr) & $\nphadralphaccpsrcost$ \\
    \npmixedlabel \,(astro) & $\npmixalphaccpostcost$ \\
    \npmixedlabel \,(psr) & $\npmixalphaccpsrcost$ \\
    \splabel \,(astro) & $\spalphaccpostcost$ \\
    \splabel \,(psr) & $\spalphaccpsrcost$ \\
    \pplabel \,(astro) & $\ppalphaccpostcost$ \\
    \pplabel \,(psr) & $\ppalphaccpsrcost$ \\
    
    \hline
    
     \end{tabular}
    \caption{Table $\chi^2/N_{\dof}$ for the $\alpha_c-C$  relations for several distributions on the EoS.  The quality of the fit decreases for all distributions except the piecewise-polytrope upon incorporating more astrophysical data, unlike the bulk of all the relations we study.}
    \label{tab:alphac-c-costs}
\end{table}

We show in  Table~\ref{tab:alphac-c-costs} the losses for this relation for all of the distributions studied.  This relation, like all others studied, shows higher losses than the I--Love--Q relations. Also similar to other relations, the nonparametric distributions show higher losses than the parametric distributions, typically by orders of magnitude.  Likewise the hadronic nonparametric distributions show  improvements in loss compared to the mixed distributions, though effects are  less than an order of magnitude

 In contrast to the other relations, however, the $\alpha_c$--$C$ relations show losses greater than one for the nonparametric EoS distributions.  This indicates that systematic errors are likely greater than statistical uncertainties for this relation.  Additionally the large errors for the piecewise-polytrope and spectral distributions  relative to other relations demonstrate that, even for these distributions, the EoS independence is questionable.  The tolerance factor we use is conservative, though removing the component which models X-ray mass-radius measurability still gives loss values greater than one, which indicates this relation very poorly models the nonparametric EoS distributions even for just the purposes of GW observations.  

 The appearance of EoS independence in, e.g., the spectral model, even though it is weak, is likely due to model-dependent correlations.  Under the spectral distribution, strong correlations appear between density scales which can lead to, e.g., the compactness (a function of the entire matter profile of the star) being correlated with the central pressure-energy density. These correlations are not present for the nonparametric EoS distributions, and are present to a weaker extent in the piecewise-polytropic EoS distributions.

%

\section{Discussion}
\label{sec:discussion}

In this paper, we  tested the EoS-independence of relations between NS properties under multiple EoS models, including parametric and nonparametric distributions.  In particular, we used a nonparametric EoS distribution, and evaluated the goodness-of-fit of the relations both to subsets mimicking hadronic EoSs or mixed-composition EoSs.  We found that effectively all relations are better fit by parametric models. Additionally within the nonparametric distributions, relations are better satisfied by EoSs which do not show signs of phase transitions.

The I--Love--Q relation is qualitatively better than other proposed relations, with typical loss values of $10^{-3}$ or below.  In particular, the $I$--$Q$ relation is very well fit by all EoS distributions.  This could be expected based on Ref.~\cite{Yagi:2014qua}, which indicated that the $I$--$Q$ relation should indeed be mostly EoS independent due to the near self-similarity of isodensity contours and near EoS independence of the elliptcity profile of NSs.   In fact, the best-fit relations we studied are  $I$--$Q$  relations under the spectral distributions, with prediction errors of $|\Delta I|/I \sim |\Delta Q|/Q \sim 0.001 $, in line with Refs.~\cite{Carson:2019rjx, Yagi:2013awa}.    
The piecewise-polytrope and nonparametric distributions are worse fit, especially for relations involving $\Lambda$.  Nonetheless even the worst-fit relation,  $Q(\Lambda)$, still has prediction errors at percent level ($\Delta Q/Q \lesssim 0.1$).  For the piecewise-polytrope model, this is qualitatively similar to the findings of Ref.~\cite{Benitez:2020fup}.   Systematic errors of $\sim 1-10\%$ are comparable to systematic errors from many other factors, such as detector calibration~\cite{Sun:2020wke}, and waveform modeling~\cite{Kunert:2021hgm, Gamba:2020wgg, Huang:2020pba, Chatziioannou:2021tdi, Read:2023hkv}.

At a comparable precision to the errors presented here, the quality of  numerical solutions to the TOV equations may become important for stars containing sharp phase transitions~\cite{Takatsy:2020bnx}.  Because the speed of sound of Gaussian process draws is analytically greater than zero, we are not subject to this concern, we do not have truly sharp transitions, and therefore standard techniques for computing the tidal Love numbers is sufficient.  Nonetheless, improved accuracy in TOV solutions will likely be important in future analyses with much improved detector sensitivities.

In contrast, the other relations involving tidal deformability  show worse fits, especially among nonparametric EoSs not informed by all astrophysical measurements. All of the C--Love, Binary-Love, and $R_{1.4}$--$\tilde \Lambda$ relations show losses of order $10^{-1}$ or more for the mixed-composition nonparametric EoS distribution.  This indicates systematic errors from these relations are already of order the statistical uncertainties. All relations, though, do improve with the inclusion of additional astrophysical data, which indicates that data have ruled out some EoS candidates inconsistent with the relations posed.  

In fitting the Binary-Love relation, the inclusion of phase transition EoSs appreciably worsens the fit to the nonparametric EoS distribution, increasing losses by a factor of 10 in the astrophysically-informed case.  This is consistent with  Ref.~\cite{Carson:2019rjx} which found that hybrid EoSs are poorly modeled with a Binary-Love relation.  In particular, ~\citet{Carson:2019rjx} found that hybrid EoSs would likely have residuals of order $\Lambda_a \sim 50$ at $\Lambda_s \sim 100 $, which is consistent with the worst-case residuals we find in Fig.~\ref{fig:binary-love-relations}.  However, the mixed-composition distributions are not universally worse-fit among relations, the C--Love fit sees comparable losses among the two distributions, indicating this relation is essentially insensitive to the presence of a phase transition.

On the other hand, the $\alpha_c$--$C$ relation is the only relation we studied with loss values greater than 1 for the nonparametric EoS distribution. A similar near total-loss of universality was observed for modes in hybrid stars~\cite{Ranea-Sandoval:2023ixr}, which could be a useful target for future work.  The loss values for the nonparametric distribution are almost 100 times worse for the nonparametric distributions than for the spectral distributions, indicating that modeling systematics are likely responsible for the appearance of EoS independence in this relation.  Nonetheless, the improvement of EoS independence in the  hadronic nonparametric case, especially upon the inclusion of additional astrophysical data, may indicate that this relation does hold universally for certain classes of EoSs (e.g. hadronic EoSs), under certain assumptions (such as astrophysically reasonable compactness-mass-radii) relations.  For this reason, even relations which are not truly EoS independent may still be useful, depending on the use cases intended.   

 \red{The goodness-of-fit improvement seen when using parametric models rather than nonparametric models is not surprising.  The parametric models have fixed functional forms which forces consistency across EoS samples within each of these sets.  Contrarily, the nonparametric EoS distribution produces EoSs with no fixed functional form and therefore no guarantee of displaying any particular phenomenology.  Therefore, we expect a much larger variety of EoS behaviors from the nonparametric distribution compared to the parametric distributions. }

These results are all dependent on the choice of tolerance factor; it is difficult to chose a completely realistic representation when many different potential sources of NS measurements exist. Nonetheless, certain conclusions, such as the relatively poor fits to the nonparametric mixed distribution relative to the spectral distribution, are independent of choice of tolerance factor. Additionally the distribution of points (NSs) that the relations are evaluated with cannot be prescribed universally.  
A potentially more physical choice that uniform-in-central-density would be a distribution which is consistent with the known population of NS sources:
\begin{equation}
    \chi^2 = \int P(\epsilon) \pi(m) \chi^2(G; F, \epsilon)dm d\epsilon,
\end{equation}
where $
\pi(m)$ is the distribution of NS masses, and $F$ is a generic NS property which serves as the independent variable for a relation and $G$ is the dependent variable of the relation.  A mapping from $F(m)$, $G(m)$ must be chosen in the case that EoSs with multiple stable branches in the $M$--$R$ relation are used. Then the loss would be equal to the expected failure of the EoS-independent relation to correctly model the next NS source detected.  However, the population of NSs observable via GWs is still poorly known~\cite{KAGRA:2021duu, Landry:2021hvl}.  Mathematically, such modifications to the analysis are equivalent to changes to the tolerance factor, though they have different physical interpretation.

It is important to recognize the sensitivity of the loss to choices such as the distribution of NSs used in evaluating each EoS $\chi^2$ and in the tolerance factor chosen for each NS.  As seen in, e.g., Fig.~\ref{fig:binary-love-relations}, the highest $\chi^2$ contributions appear at high $\Lambda$ values for relations involving $\Lambda$, equivalent to larger residuals there (under the constant uncertainty model).  There may not exist merging BNSs with symmetric tidal deformabilities as high as $10^4$, or they may be exceedingly rare.  However, Fig.~\ref{fig:binary-love-relations} also demonstrates that at $\Lambda \sim 10^3$, deviations from the EoS-independent relation of order 100 or larger are still possible within the nonparametric model.  Therefore, we expect variation in the loss based on choices in the truncation of the population, though we do not expect the relationship between losses for the various models to change appreciably under different assumptions. Additionally, assessing the EoS independence of relations when matter is not in cold $\beta$-equilibrium, or when NSs are not isolated and nonspinning~\cite{Largani:2021hjo}, may be challenging.  In particular, NS merger remnants may be highly spinning, hot, and dynamically perturbed, so the cold relations explored here, and the strategy used to evaluate them, will likely have to be extended.  
Longer-term EoS independence tests will likely have to carefully examine all of these factors in order to determine, with higher fidelity, the usefulness of EoS-independent relations to our understanding of NSs and the nuclear EoS.

\section{Acknowledgments}

We thank Victor Guedes for code to compute the quadrupole moment for NSs.
We thank Lami Suleiman for comments on the manuscript.
This work is supported by National Science Foundation grant No. PHY-2150027 as part of the LIGO Caltech REU Program.
I.L. and K.C. acknowledge support from the Department of Energy under award number DE-SC0023101. 
K.C. acknowledges support from the Sloan Foundation.
R.E. is supported by the Natural Sciences \& Engineering Research Council of Canada (NSERC).
The authors are grateful for computational resources provided by the LIGO Laboratory and supported by National Science Foundation Grants PHY-0757058 and PHY-0823459.
Software: This work makes use of \texttt{scipy}~\cite{SciPyNMeth}, \texttt{numpy}~\cite{numpy}, and \texttt{matplotlib}~\cite{Hunter:2007}.  This material is based upon work supported by NSF's LIGO Laboratory which is a major facility fully funded by the National Science Foundation.  The authors gratefully acknowledge the Italian Instituto Nazionale de Fisica Nucleare (INFN), the French Centre National de la Recherche Scientifique (CNRS), and the Netherlands Organization for Scientific Research for the construction and operation of the Virgo detector and the creation and support of the EGO consortium.

\appendix
\section{Additional Figures and Tables}
\label{app:additional-figures}

In this appendix we display results for the piecewise-polytrope and nonparametric hadronic EoS distributions, in similar form as the main text results for the spectral and nonparametric mixed distributions. The $\chi^2$ values for each fit  are given in the main text tables. We use the astrophysically-informed EoS distributions, again because we do not find significant differences modulo improvements of order no more than 10 to the fit quality upon conditioning.

\begin{figure*}[htp!]
    \centering
    \includegraphics[height=7.5cm, width=.32\textwidth]{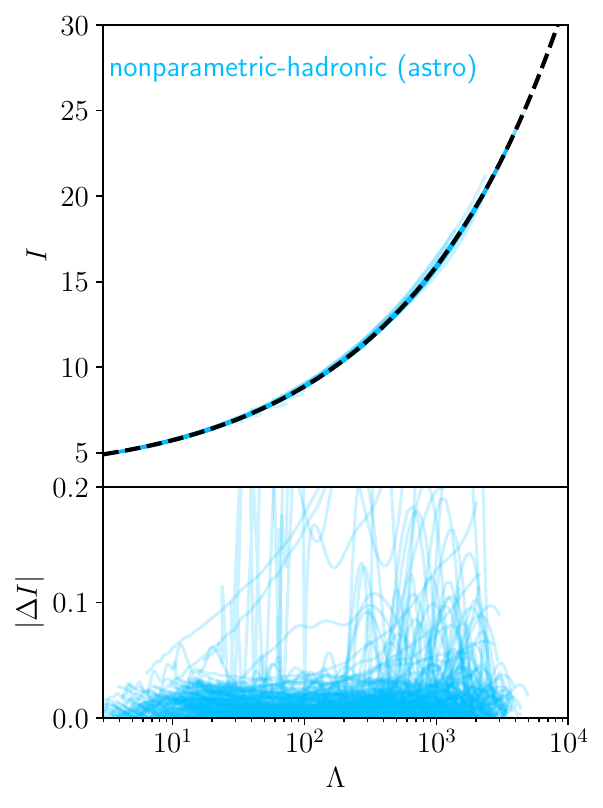}
    \includegraphics[height=7.5cm, width=.32\textwidth]{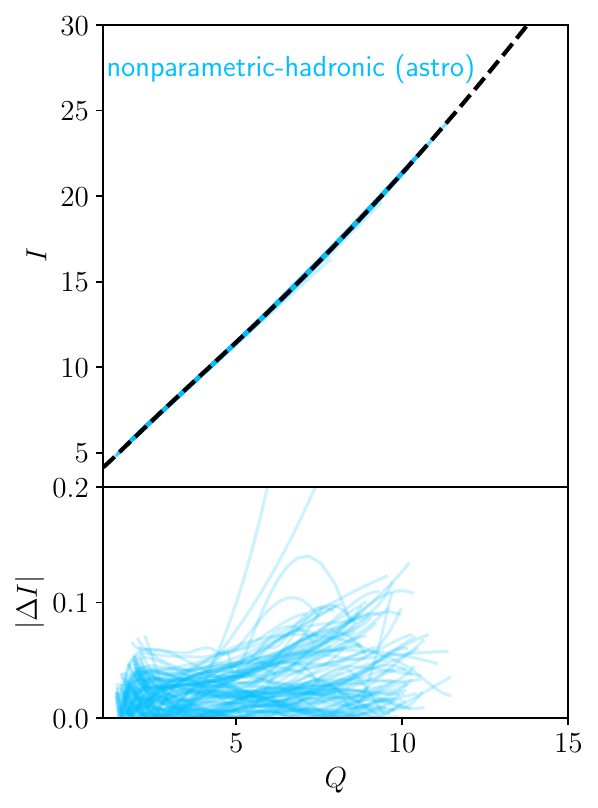}
    \includegraphics[height=7.4cm, width=.32\textwidth]{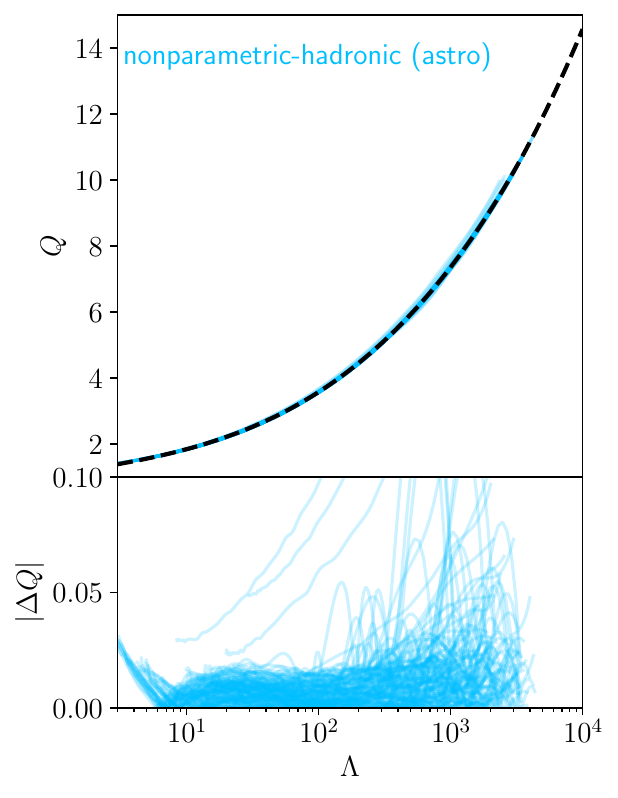}
    \caption{The same as Fig.~\ref{fig:I--Love--Q-fits-np-mixed-post}, but with the hadronic nonparametric distribution conditioned on all astrophysical data.}  
    \label{fig:I--Love--Q-fits-np-hadronic-post}
\end{figure*}

\begin{figure*}[htp!]
    \centering
    \includegraphics[height=7.5cm, width=.32\textwidth]{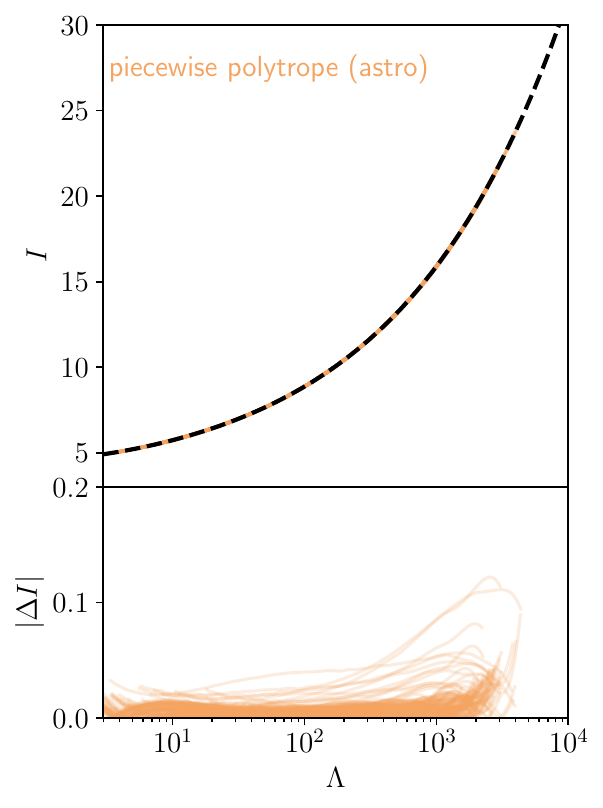}
    \includegraphics[height=7.5cm, width=.32\textwidth]{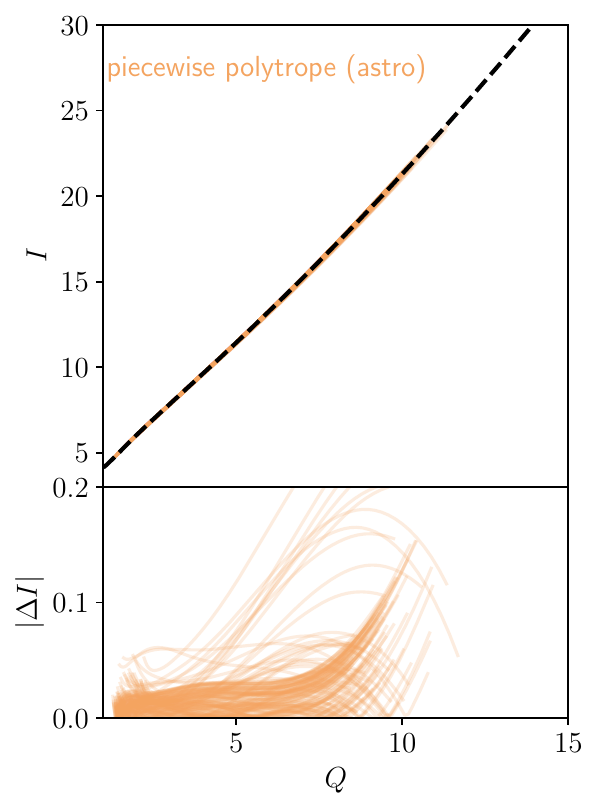}
    \includegraphics[height=7.4cm, width=.32\textwidth]{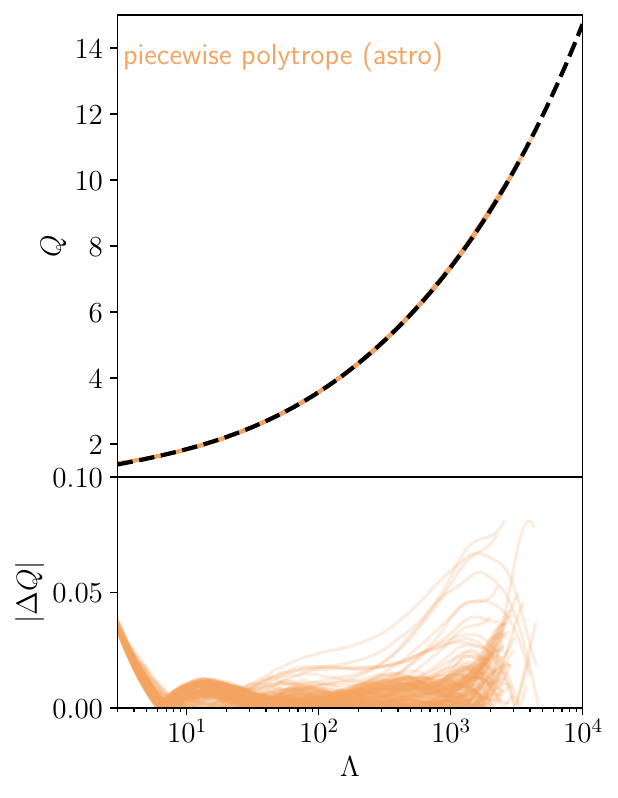}
    
    \caption{The same as Fig.~\ref{fig:I--Love--Q-fits-np-mixed-post} but with the piecewise-polytrope EoS distribution conditioned on all astrophysical data.}
    \label{fig:I--Love--Q-fits-pp-post}
\end{figure*}

For the I--Love--Q relation, we display the fits for the hadronic nonparametric distribution in Fig.~\ref{fig:I--Love--Q-fits-np-hadronic-post}, and for the piecewise-polytrope EoS distribution in Fig.~\ref{fig:I--Love--Q-fits-pp-post}.  $I$--$Q$ is still the best fit EoS-independent relation.   We also give fitting coefficients in Table~\ref{tab:I--Love--Q-coefficients-np-sp}.

\begin{table*}[]
    \centering
    \begin{tabular}{|c|}
    \hline
    \\
     Coefficient\\
     \hline
        $\alpha$ \\
         $K_{yx}$ \\
         $a_1$\\
         $a_2$\\
         $a_3$\\
         $b_1$\\
         $b_2$\\
         $b_3$ \\  
         \hline
    \end{tabular}
    \begin{tabular}{|c|c|c|}
    \hline
    \multicolumn{3}{|c|}{GP-mixed (astro)}\\
    \hline
    $I(\Lambda)$ &$Q(\Lambda)$&$I(Q)$\\
     \hline
     2/5 & 1/5 & 2\\
0.5356 & 0.0072 & 0.0072 \\
1.7583 & 11.1589 & 11.1589 \\
1.3883 & -37.6926 & -37.6926 \\
-5.6089 & 42.7718 & 42.7718 \\
-0.7071 & -2.5557 & -2.5557 \\
-0.9748 & 2.3251 & 2.3251 \\
0.5105 & -7.3937 & -7.3937 \\
\hline
     \end{tabular}
    \begin{tabular}{|c|c|c|}
    \hline
        \multicolumn{3}{|c|}{SP (astro)}\\
        \hline
    $I(\Lambda)$ &$Q(\Lambda)$&$I(Q)$\\
     \hline
      2/5 & 1/5 & 2\\
0.5139 & 0.0052 & 0.0052 \\
2.0486 & 12.1774 & 12.1774 \\
0.8249 & -37.0504 & -37.0504 \\
6.7629 & 43.0395 & 43.0395 \\
-1.0615 & -2.6161 & -2.6161 \\
2.2034 & 2.4074 & 2.4074 \\
-0.9326 & -7.6697 & -7.6697 \\
\hline
     \end{tabular}
     \begin{tabular}{|c|c|c|}
     \hline
        \multicolumn{3}{|c|}{PP (astro)}\\
        \hline
    $I(\Lambda)$ &$Q(\Lambda)$&$I(Q)$\\
     \hline
      2/5 & 1/5 & 2\\
0.4192 & 0.0007 & 0.0007 \\
2.2881 & 30.3545 & 30.3545 \\
-3.7192 & -16.5079 & -16.5079 \\
72.9633 & 48.3485 & 48.3485 \\
-3.5874 & -2.7902 & -2.7902 \\
16.8924 & 2.6775 & 2.6775 \\
-7.6431 & -8.7651 & -8.7651 \\
\hline
     \end{tabular}
    \caption{Table of coefficients for the I--Love--Q relations for the nonparametric mixed-composition, spectral, and piecewise-polytrope astrophysical posterior EoS distributions.  See Eqs.~\eqref{eq:i-love-relation},~\eqref{eq:q-love-relation}, and~\eqref{eq:i-q-relation} respectively. }
    \label{tab:I--Love--Q-coefficients-np-sp}
\end{table*}

\begin{figure*}
    \centering
    \includegraphics[width=.49\textwidth]{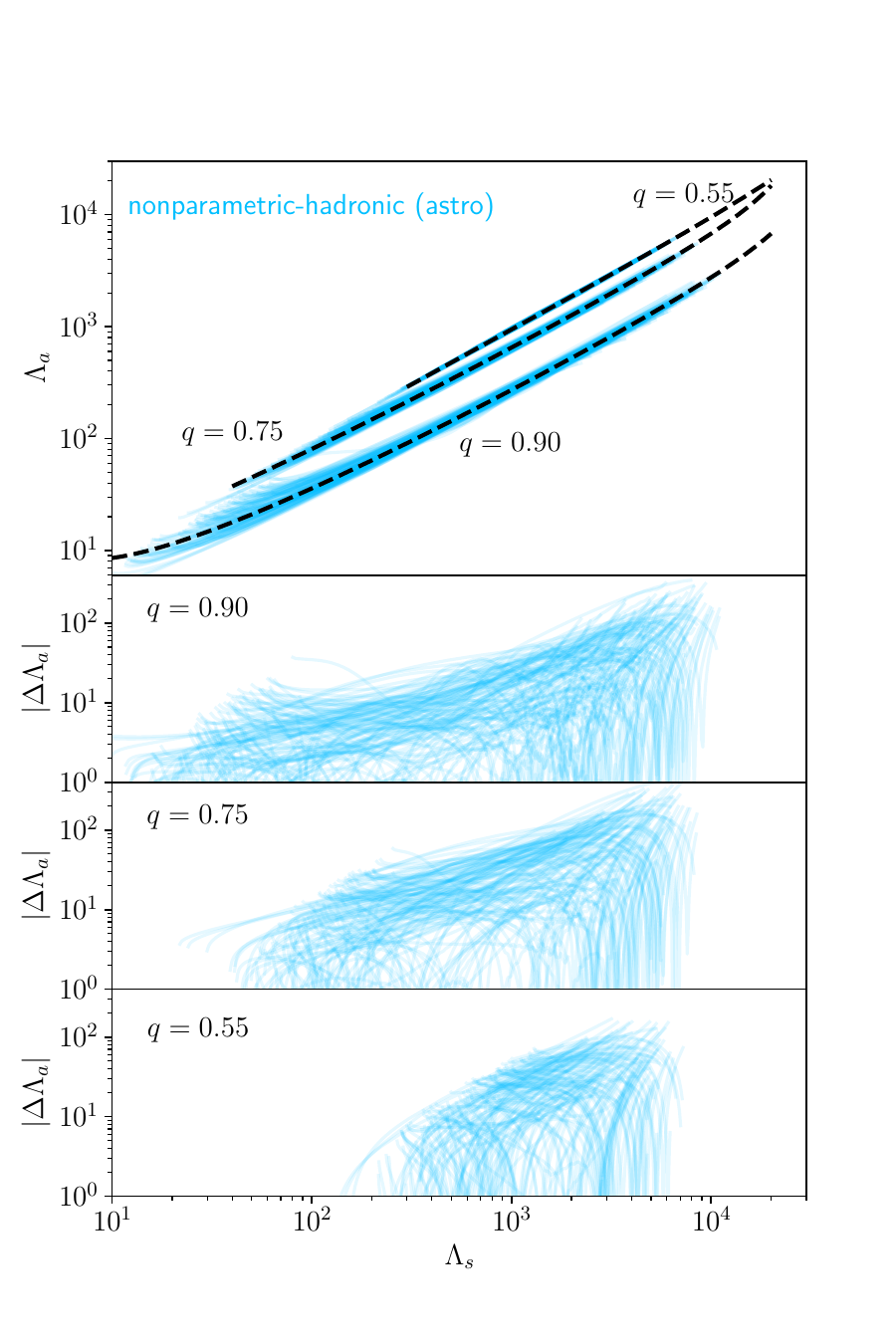}
    \includegraphics[width=.49\textwidth]{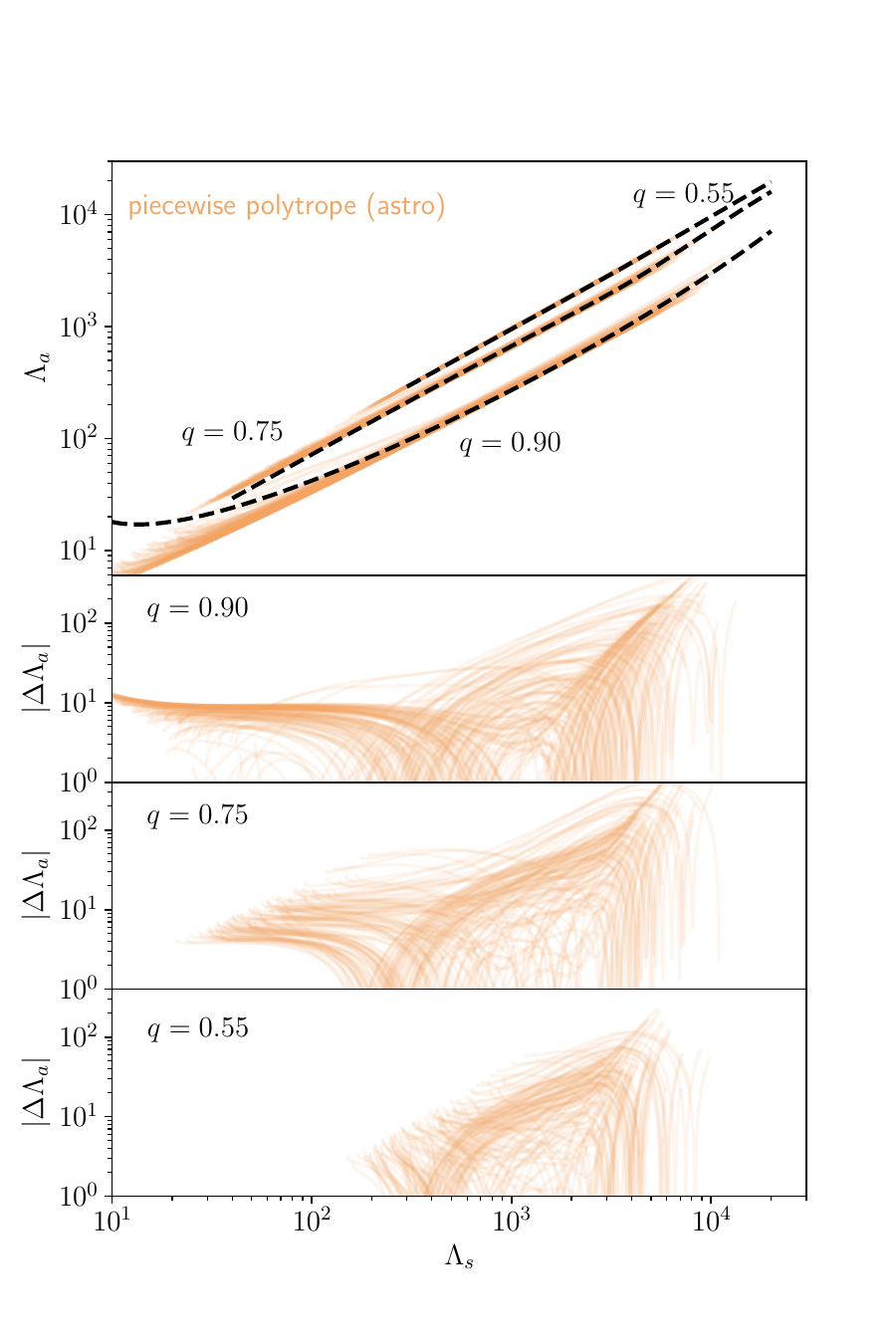}
    \caption{The same as Fig.~\ref{fig:binary-love-relations}, but with the hadronic nonparametric EoS distribution on the left, and the piecewise-polytrope distribution on the right.  }
    \label{fig:binary-love-relations-hadronic-pp}
\end{figure*}

We display the fits for the Binary-Love relation for the hadronic nonparametric distribution and piecewise-polytrope distribution in Fig.~\ref{fig:binary-love-relations-hadronic-pp}.   We display the best-fit coefficients in Table~\ref{tab:binary-love-coefficients-np-sp}.

\begin{table*}[]
    \centering

    \begin{tabular}{|c|}
        \hline
    \\
     Coefficient\\
     \hline
$b_{11}$ \\ $b_{12}$ \\ $b_{21}$ \\ $b_{22}$ \\ $b_{31}$ \\ $b_{32}$ \\ $c_{11}$ \\ $c_{12}$ \\ $c_{21}$ \\ $c_{22}$ \\ $c_{31}$ \\ $c_{32}$\\
\hline
    \end{tabular}
    \begin{tabular}{|c|c|c|}
    \hline
    \multicolumn{3}{|c|}{GP-mixed (astro)}\\
    \hline
    $q=.55$ &$q=.75$&$q=.9$\\
     \hline
  -13.6363 & -13.0286 & -35.8727 \\
16.6082 & 11.9352 & 38.1073 \\
60.9451 & 49.595 & 17.3353 \\
-22.4131 & -15.7433 & -10.7606 \\
-132.7392 & -95.409 & -85.2499 \\
-35.1957 & -33.9646 & 66.3908 \\
-36.1830 & -121.3958 & 20.9222 \\
62.3338 & 157.5113 & -24.9027 \\
60.2142 & 57.6574 & 44.7671 \\
-27.5988 & -38.6213 & -42.4961 \\
-132.3099 & -89.4668 & -25.8278 \\
-18.1968 & -6.9804 & 4.5299 \\
\hline
     \end{tabular}
         \begin{tabular}{|c|c|c|}
         \hline
        \multicolumn{3}{|c|}{SP (astro)}\\
        \hline
    $q=.55$ &$q=.75$&$q=.9$\\
     \hline
89.3902 & -114.1202 & -13.7126 \\
-199.4862 & 153.6543 & 14.5353 \\
137.5437 & 65.0962 & 30.3579 \\
-97.1215 & -86.1626 & -36.4909 \\
-227.1308 & -150.6500 & -21.6415 \\
-12.4604 & -37.6394 & 20.1071 \\
-34.3331 & -17.6402 & -1.2833 \\
49.3492 & 30.3484 & 0.9211 \\
69.5378 & -46.8838 & 34.3594 \\
24.7469 & 2.4981 & -43.3557 \\
-193.2270 & -145.6506 & -29.3931 \\
-43.5078 & 142.7302 & 36.7860 \\ 
\hline
 \end{tabular}
\begin{tabular}{|c|c|c|}
    \hline
        \multicolumn{3}{|c|}{PP (astro)}\\
        \hline
    $q=.55$ &$q=.75$&$q=.9$\\
     \hline
-13.318 & -16.0181 & -14.2241 \\
14.5180 & 13.5785 & 14.6087 \\
60.5649 & 60.5607 & 29.9721 \\
-17.0596 & -9.8172 & -33.4993 \\
-125.4819 & -117.0661 & -22.0825 \\
-31.7949 & -50.6147 & 20.6718 \\
-27.6276 & -79.3091 & -14.3035 \\
43.9329 & 99.8593 & 15.0504 \\
63.5022 & 62.0146 & 35.7453 \\
-30.8961 & -31.4480 & -43.0991 \\
-125.3372 & -101.7505 & -28.5082 \\
-11.8738 & -18.4297 & 35.8010 \\
\hline

     \end{tabular}
    \caption{Table of coefficients for the Binary-Love relations for the nonparametric mixed-composition,  spectral, and piecewise-polytrope posterior EoS distributions.  See Eq.~\eqref{eq:binary-love-relation}.}
    \label{tab:binary-love-coefficients-np-sp}
\end{table*}

\begin{figure*}
    \centering
    \includegraphics[width=.49\textwidth]{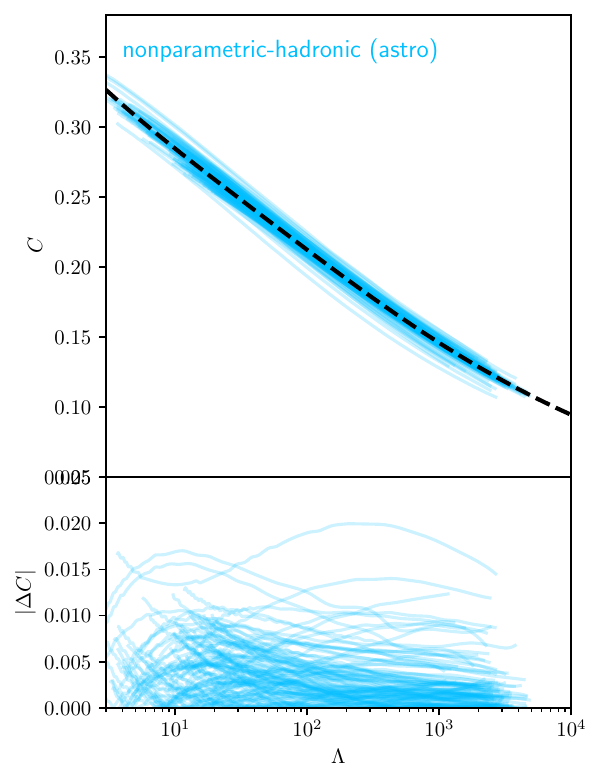}
    \includegraphics[width=.49\textwidth]{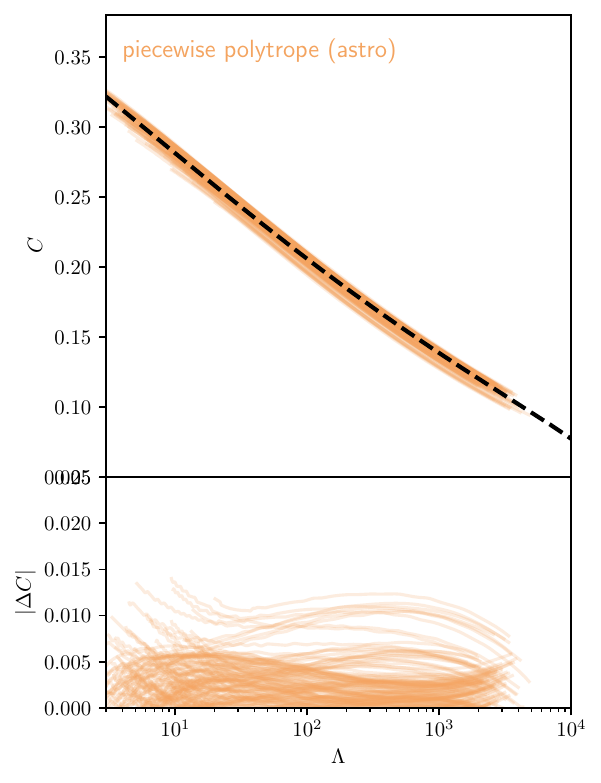}
    \caption{The same as Fig.~\ref{fig:C-Love-relations}, but with the hadronic nonparametric EoS distribution on the left, and the piecewise-polytrope EoS distribution on the right.  Both 
    distributions are conditioned on all astrophysical data.}
    \label{fig:C-Love-relations-hadronic-pp}
\end{figure*}

We display the fits for the C--Love relation for the hadronic nonparametric distribution and piecewise-polytrope distribution in Fig.~\ref{fig:C-Love-relations-hadronic-pp}. We display the best-fit coefficients in Table~\ref{tab:C-Love-coefficients-np-sp}.

\begin{table}[]
\centering
\begin{tabular}{|c|}
\hline
\\

 Coefficient\\
 \hline
     $K_{yx}$ \\
     $a_1$\\
     $a_2$\\
     $a_3$\\
     $b_1$\\
     $b_2$\\
     $b_3$\\
     \hline
\end{tabular}
\begin{tabular}{|c|}
\hline
\multicolumn{1}{|c|}{GP-mixed (astro)}\\
\hline
$C(\Lambda)$\\
 \hline
0.0833 \\
-529.6368 \\
666.1701 \\
-1119.5632 \\
-84.2438 \\
144.0589 \\
-2.7723 \\
\hline
 \end{tabular}
     \begin{tabular}{|c|}
     \hline
    \multicolumn{1}{|c|}{SP (astro)}\\
    \hline
$C(\Lambda)$\\
 \hline
1.9392 \\
-96.4366 \\
-69.8059 \\
-191.0251 \\
-360.1569 \\
152.5207 \\
-1702.2789 \\
\hline
 \end{tabular}
  \begin{tabular}{|c|}
  \hline
    \multicolumn{1}{|c|}{PP (astro)}\\
\hline  
$C(\Lambda)$\\
 \hline
3.5446 \\
-28.1750 \\
-127.7955 \\
-43.2623 \\
-191.1053 \\
-433.2343 \\
-1318.6131 \\
\hline
 \end{tabular}
\caption{Table of coefficients for the C--Love relations for the nonparametric mixed-composition, spectral, and piecewise-polytrope astrophysical posterior EoS distributions.  See Eq.~\eqref{eq:c-lambda-relation}.}

\label{tab:C-Love-coefficients-np-sp}
\end{table}

For the $\tilde \Lambda$--$R_{1.4}$ relation we display the value for the coefficient $a$  for all of the EoS sets in Table~\ref{tab:rtyp-lambdatilde-coefficients-np-sp}.
\begin{table}[]
    \centering
    \begin{tabular}{|c|c|c|c|}
    \hline  
     Coefficient & GP-mixed (astro) & SP (astro) & PP (astro) \\
     \hline
    $a$   & 3.6387 &   3.7867 & 3.8086\\
    \hline
     \end{tabular}
    \caption{Table of coefficients for the $R_{1.4}$--$\tilde \Lambda$ relation.  See Eq.~\eqref{eq:r1p4-lambda-tilde}.  }
    \label{tab:rtyp-lambdatilde-coefficients-np-sp}
\end{table}

\begin{figure*}
    \centering
    \includegraphics[width=.49\textwidth]{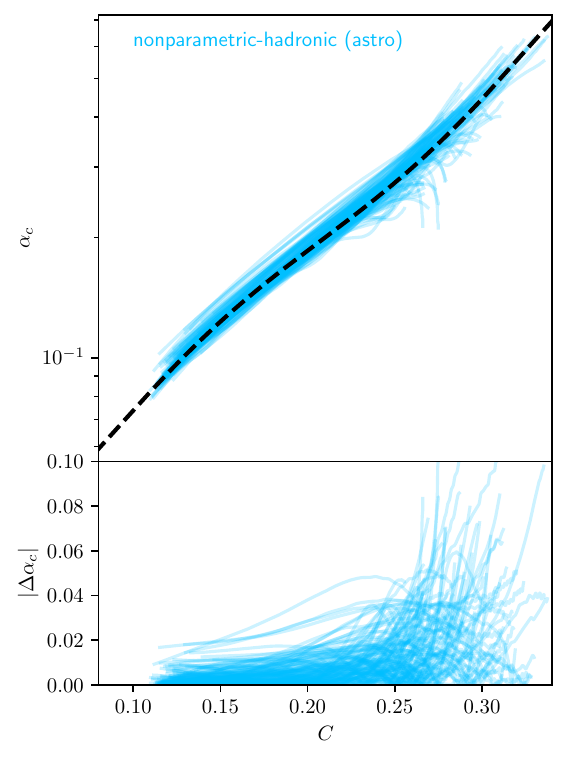}
    \includegraphics[width=.49\textwidth]{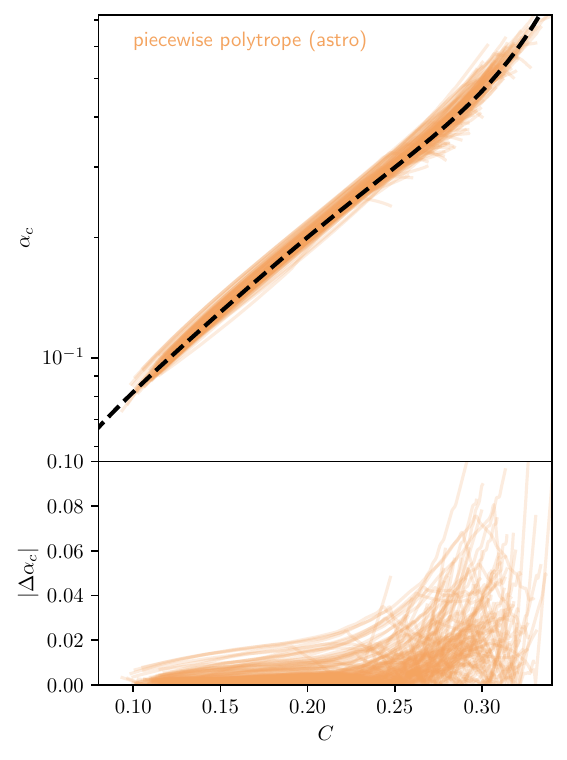}
    \caption{Left: The same for the nonparametric EoS distribution, conditioned on all astrophysical data.
    Right : The same for the piecewise-polytrope parametrization, conditioned on all astrophysical data. Same as Fig.~\ref{fig:alphac-c-relations}.}
    \label{fig:alphac-c-relations-hadronic-pp}
\end{figure*}

We display the fits to the $\alpha_c$--$C$ EoS-independent relation for the nonparametric hadronic, and piecewise-polytrope distribution both conditioned only on mass measurements of heavy pulsars,  and conditioned on all astrophysical data in Fig.~\ref{fig:alphac-c-relations-hadronic-pp}.  The piecewise-polytropic distribution is the only one which is better fit by the $\alpha_c$--$C$ relation after the inclusion of GW mass-tidal deformability and X-ray mass-radius measurements.  This can be attributed to \emph{a priori} large values of $\alpha_c$ in the cores of the most massive neutron stars under the piecewise-polytrope models.     
\begin{table}[]
    \centering
    \begin{tabular}{|c|}
    \hline
    \\
     Coefficient\\
     \hline
        $a_0$\\
         $a_1$\\
         $a_2$\\
         $a_3$\\
         $a_4$\\
         $a_5$\\
         \hline
    \end{tabular}
    \begin{tabular}{|c|}
    \hline
    \multicolumn{1}{|c|}{GP-mixed (astro)}\\
    \hline
    $\alpha_c(C)$\\
     \hline
-7.3477 \\
88.5223 \\
-591.4298 \\
1960.4713 \\
-2799.1485 \\
1215.7415 \\
\hline
     \end{tabular}
         \begin{tabular}{|c|}
         \hline
        \multicolumn{1}{|c|}{SP (astro)}\\
        \hline
    $\alpha_c(C)$\\
     \hline
-5.1067 \\
49.9461 \\
-379.9054 \\
1729.6551 \\
-3988.1116 \\
3783.0214 \\
\hline
     \end{tabular}
      \begin{tabular}{|c|}
      \hline
        \multicolumn{1}{|c|}{PP (astro)}\\
        \hline
    $\alpha_c(C)$\\
     \hline
-4.7738 \\
45.8993 \\
-389.7208 \\
2051.4967 \\
-5396.4668 \\
5.6082 \\
\hline

     \end{tabular}
    \caption{Table of coefficients for the $\alpha_c$--$C$ relation for the nonparametric mixed-composition and Spectral posterior EoS distributions. See Eq.\eqref{eq:c-alphac-relation}.}
    \label{tab:c-alphac-coefficients-np-sp}.
\end{table}

\bibliography{References.bib}
\end{document}